\documentclass[usenatbib]{mnras}
\usepackage{amsmath}	% Advanced maths commands

\usepackage[T1]{fontenc}
\usepackage{ae,aecompl}

\usepackage{graphicx}	% Including figure files
\usepackage{amssymb}	% Extra maths symbols

\newcommand{\lele}[3]{{#1}\,$\le$\,{#2}\,$\le$\,{#3}}

\newcommand{\tiunit}{$J\,m^{-2}\,s^{-1/2}\,K^{-1}$}

\begin{document}

\renewcommand{\theenumi}{\alph{enumi}}
\def\gv{2002\,GV\ensuremath{_{31}}}
\def\wg{2010\,WG\ensuremath{_{9}}}
\def\jj{(278331)\,2007\,JJ\ensuremath{_{43}}}
\def\jjs{2007\,JJ\ensuremath{_{43}}}
\def\ortiz{2007\,OR\ensuremath{_{10}}}
\def\ortizlong{(225088) 2007\,OR\ensuremath{_{10}}}
\sloppy

\title[Nereid from space]{Nereid from space: 
Rotation, size and shape analysis from Kepler/K2, Herschel and Spitzer observations }

\author[Cs. Kiss et al.]{
Cs.~Kiss,$^{1}$
A.~P\'al,$^{1,2}$
A.~I.~Farkas-Tak\'acs$^{1,2}$
Gy.~M.~Szab\'o,$^{1,3,4}$
R.~Szab\'o,$^{1}$
\and
L.~L.~Kiss,$^{1,4,5}$
L.~Moln\'ar,$^{1}$
K.~S\'arneczky,$^{1,4}$
Th.G.~M\"uller,$^{6}$
M.~Mommert,$^{7}$
\and and J.~Stansberry$^{8}$\\
$^{1}$Konkoly Observatory, Research Centre for Astronomy and Earth Sciences, Hungarian 
Academy of Sciences\\
H-1121 Budapest, Konkoly Thege Mikl\'os \'ut 15-17, Hungary\\
$^{2}${E\"otv\"os Lor\'and Tudom\'anyegyetem, H-1117 P\'azm\'any P\'eter s\'et\'any 
1/A, Budapest, Hungary}\\
$^{3}${ELTE Gothard Astrophysical Observatory, H-9704 Szombathely, 
Szent Imre herceg \'ut 112, Hungary}\\
$^{4}${Gothard-Lend\"ulet Research Group, H-9704 Szombathely, Szent Imre 
herceg \'ut 112, Hungary}\\
$^{5}${Sydney Institute for Astronomy, School of Physics A28, University of Sydney, 
NSW 2006, Australia}\\
$^{6}$Max-Planck-Institut f\"ur extraterrestrische Physik, Postfach 1312, Giessenbachstr.,
85741 Garching, Germany\\
$^{7}$Department of Physics and Astronomy, Northern Arizona University, P.O. Box 6010, 
Flagstaff, AZ 86011, USA\\
$^{8}$Space Telescope Science Institute, 3700 San Martin Dr., Baltimore, MD 21218, United States}

%\begin{document}

\label{firstpage}
\pagerange{\pageref{firstpage}--\pageref{lastpage}}
\date{\today}
\maketitle

\begin{abstract}
In this paper we present an analysis of Kepler K2 mission Campaign~3 
observations of the irregular Neptune satellite, Nereid. We determined a rotation period 
of P\,=\,11.594$\pm$0.017\,h and amplitude of $\Delta$m\,=\,0\fm0328$\pm$0\fm0018,
confirming previous short rotation periods obtained in ground based observations. 
The similarities of light curve amplitudes between 2001 and 2015 show that Nereid is 
in a low-amplitude rotation state nowadays and it could have been
in a high-amplitude rotation state in the mid 1960's. Another high-amplitude
period is expected in about 30 years. Based on the light curve amplitudes observed
in the last 15 years we could constrain the shape of Nereid and obtained 
a maximum $a$:$c$ axis ratio of 1.3:1. This excludes the previously suggested
very elongated shape of $a$:$c$\,$\approx$\,1.9:1 and clearly shows that Nereid's
spin axis cannot be in forced precession due to tidal forces. Thermal emission data
from the Spitzer Space Telescope and the Herschel Space Observatory indicate that 
Nereid's shape is actually close to the $a$:$c$ axis ratio limit of 1.3:1 we obtained, 
and it has a very rough, highly cratered surface.

\end{abstract}

\begin{keywords}
methods: observational --- 
		techniques: photometric --- 
		astrometry --- 
		planets and satellites: individual: Neptune II Nereid 		
\end{keywords}

\section{Introduction}
\label{sect:introduction}
Nereid is a large, $\sim$350\,km sized, 
irregular satellite orbiting Neptune in a very eccentric and inclined 
orbit \citep{Dobrovolskis,Jacobson2009}. 
Although its orbital characteristics are well known, there is still 
a mistery related to its shape, orientation and rotation rate. 
There have been several papers reporting on unusual, in some cases large amplitude 
brightness variations of Nereid on different timescales, from night-to-night
variations to changes on annual scales. These investigations are nicely summarised and discussed 
in \citet{Schaefer2008}. In this paper the authors' preferred solution that could
explain the photometric variability was that year-to-year variations are caused by
Nereid's pole precessing, in some years Nereid's pole pointing towards Earth while in other years 
nearly perpendicular to the line of sight. Physically this could be explained by 
forced precession, due to Neptune's tidal torque on a non-spherical
Nereid. This would require high, 1.9:1 or greater axis ratios. 

\citet{Hesselbrock2013} attempted to model the rotation of Nereid in a way similar to 
\citet{Schaefer2008}, but also considering
the effect of smaller bodies in addition to the Sun and Neptune, especially that of Triton. 
Their "best estimate" solution is a triaxial ellipsoid with semiaxis ratios of 
c/a$\sim$0.5, b/a$\sim$0.6, an inital obliquity of $\sim$60\degr{} and an initial 
rotation period of 144\,h. Their model predicts "active" and "inactive" periods, depending
on the actual direction of the main spin axis 
(c-axis aspect angles of $\sim$90\degr{} and $\sim$0\degr, respectively), matching the 
photometry data collected in \citet{Schaefer2008} qualitatively, however, with some
differences in the predicted brightness levels compared with the observed ones. 
They explain this latter discrepancy by an additional effect of a non-uniform albedo
distribution. Their model predicts a $\sim$15-year time between the two extreme orientations.  

While the shape of Nereid would be essential for the rotation evolution
models, this is not known. \citet{Thomas1991} derived a size of 350$\pm$50\,km from
Voyager-2 flyby observations in which Nereid was resolved to some level, 
but the quality of the data  was not good enough to constrain the shape 
\citep[see also][for a discussion]{Schaefer2008}. 

Based on observations in August 2001 and August 2002 \citet{Grav2003} derived 
a short rotation period of P\,=\,11\fh52 for Nereid with a small peak-to-peak 
light curve amplitude of 0\fm029$\pm$0.003. This rotation period was later questionned 
by \citet{Schaefer2008} due to the relatively poor sampling of the rotation curve, 
and has not been included in the analysis by \citet{Hesselbrock2013} either.  
\citet{Terai} obtained a very similar light curve 
($\Delta$m\,=\,0\fm031$\pm$0\fm001, P\,=\,11\fh50$\pm$0\fh10)
based on observations in 2008, but their light curve was sparsely 
sampled in terms of rotational phase. 

As shown above, ground based observations have not placed a strong constraint on
the rotation of Nereid. However, as it has been demonstrated in some recent 
papers \citep{pal2015,pal2015c}, data from
the extended Kepler mission \citep[K2,][]{howell2014} 
can be very effectively used to obtain rotational light
curves of distant Solar System bodies due to uninterrupted photometric time series
of several tens of days in length. Given that Neptune, and therefore Nereid as well, was
included in the Campaign~3 observations of the K2 mission, we attempted to obtain 
the light curve of Nereid from these observations. 
In addition to the light curve data, we collected archival Spitzer Space 
Telescope and Herschel Space Observatory data to detect the 
thermal emission of Nereid in the mid- and far-infrared (24--160\,$\mu$m)
wavelength ranges. In this paper, with the synergy of the light curve information 
and thermal emission data, we put important constraints on the rotational state, 
shape, as well as on other physical properties of Nereid.   

%%%%%%%%%%%%%%%%%%%%%%%%%%%%%%%%%%%%%%%%%%%%%%%%%%%%%%%%%%%%%%
%\section{Observations and data reduction}

%%%%%%%%%%%%%%%%%%%%%%%%%%%%%%%%%%%%%%%%%%%%%%%%%%%%%%%%%%%%%%%%
\section{The Kepler K2 light curve}

\subsection{K2 observations}
\label{sect:observations}

{\it Kepler} observed Nereid during the third campaign of the extended mission,
named K2 \citep{howell2014}. In this K2 mission,
{\it Kepler} targets fields near the Ecliptic, observing each field for
approximately three months. This quarterly schedule allows the continuous
observations of Solar System bodies if sufficiently large arcs are allocated
in the CCD mosaic of the {\it Kepler} space telescope. Main-belt asteroids
have a large apparent motion (comparable to or even larger than the total 
field-of-view of the space telescope). In addition, main-belt asteroids have a 
non-negligible noise source on the photometry of stellar 
targets \citep{szabo2015}. However, minor bodies outside the main-belt
-- such as Centaurs and trans-Neptunian objects -- can be observed with a 
relatively low pixel budget. In addition, the pixel budget can even be 
minimized around the stationary point, where masks with a size of 
$\sim20\times20$ pixels can be sufficient to follow trans-Neptunian objects
up to $10-15$ days \citep{pal2015}. 

%% %% %% %% %% %% %% %% %% %% %% %% %% %% %% %% %% %% %% %% %% %% %% %% %% %% 
\begin{figure}
\resizebox{80mm}{!}{\includegraphics{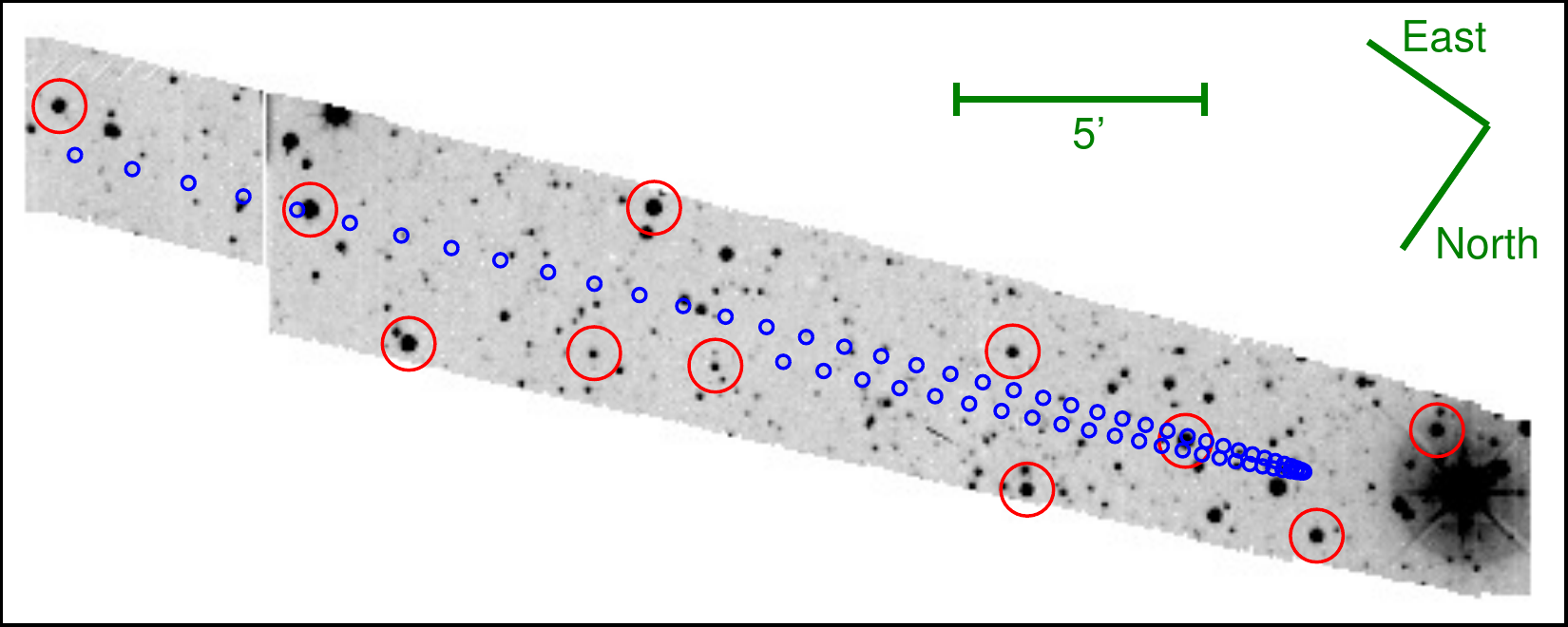}}
\caption{
The field-of-view of {\it Kepler}, showing the field 
in which Neptune and Nereid was apparently moving during the K2 
Campaign 3. The stars involved in the determination of the absolute
and differential astrometric solutions are indicated by red circles.
The small blue circles indicate the position of Nereid with a 
$1$-day stepsize throughout the observations.
The total size of the field is $475\times190$ pixels, i.e. 
$\sim 32^\prime\times13^\prime$. This image is shown in the CCD frame,
hence the image is flipped with respect to the standard view.
}
\label{fig:nereidfield}
\end{figure}
%% %% %% %% %% %% %% %% %% %% %% %% %% %% %% %% %% %% %% %% %% %% %% %% %% %% 

In the case of Nereid, a nearly parallelogram-shaped field was allocated for
{\it Kepler}/K2 observations. This field includes the apparent track of
both Neptune and Nereid and has a size of $456\times 80$ pixels.
This field is located on {\it Kepler} module \#13 and read out by 
its output amplifier \#2 (i.e. channel \#42 considering the full mosaic array).
In order to analyze the field, we retrieved the $456$ individual data 
frames corresponding to the columns of this parallelogram-shaped CCD area.
Thereafter, we built the individual image frames on a rectangular
area having a size of $475\times190$. This area safely covers the 
parallelogram and includes some additional rows and columns in order to have
same extra space for frame registration. In total, the corresponding data
series contains $3336$ individual and useable long-cadence frames. The total
number of frames for Campaign 3 were $3386$, but $50$ frames had to be
dropped due to lack of data or inappropriate tracking during the exposure.

Due to the lack of a third active reaction wheel, the positioning jitter of
{\it Kepler}/K2 is significantly larger than what it had been during the main
mission. Therefore, the $3336$ frames have to be analyzed independently
in order to retrieve the proper astrometric solution as well as to
perform an image registration needed by the subsequent steps of differential
photometry. To accomplish this astrometry, we selected nearly a dozen 
reference stars distributed uniformly in the field. Due to the large size
of the field, we did not include additional stamps for this purpose
\citep[see e.g. the case of \ortiz{},][]{pal2015c}. These stars were
used to perform both the differential astrometry (i.e. the frame registration)
and the absolute astrometry (i.e. finding the plate solution w.r.t.
the J2000 system). The absolute plate solution has been derived by cross-matching
the pixel coordinates of the selected reference stars with the
USNO-B catalogue \citep{monet2003}.
These $11$ stars were also used to 
transform {\it Kepler}/K2 photometry to USNO-B1 R system. The USNO magnitudes
of these star cover the range of $R=13.2\dots17.8$, nearly homogeneously.
Since an unfiltered CCD efficiency curve can be considered as a ``wide R''
band, the comparison of the USNO magnitudes with the instrumental ones
yielded a fit residual of $0.12$ magnitudes. Therefore, we can state that the
accuracy of this transformation is in this range.

In all other aspects, the data reduction and photometry were conducted
in a similar manner as it was described in \cite{pal2015} or \cite{pal2015c}.
Our photometric pipeline used for {\it Kepler} frame reconstruction, source
extraction, astrometry and cross-matching, image registration and
differential photometry are based on the various tasks of the FITSH
package \citep{pal2012}. The folded light curve (with the period of 
$11.594\,{\rm h}$, see Sect.~\ref{sect:analysis}) of Nereid is shown 
in Fig.~\ref{fig:lc}.
We note here that the formal photometric 
uncertainities on the individual frames increased from $0.012$ up to 
$0.022$ throughout this $\sim 67$\,days long campaign. This increment is 
due to the gradually increasing level of zodiacal light as the elongation of 
Nereid decreased from $140^\circ$ down to $74^\circ$. This instrumental 
uncertainity can be compared with the statistical one computed from the 
root mean square deviation of the photometric data points on each of the $N=20$
bins on the binned light curve. This latter method gives us an estimation
of $0.028-0.034$ error for each frame, i.e. slightly larger than the 
instrumental estimate.

\subsection{Light curve analysis}
\label{sect:analysis}

\begin{figure}
\centering\includegraphics[width=0.45\textwidth]{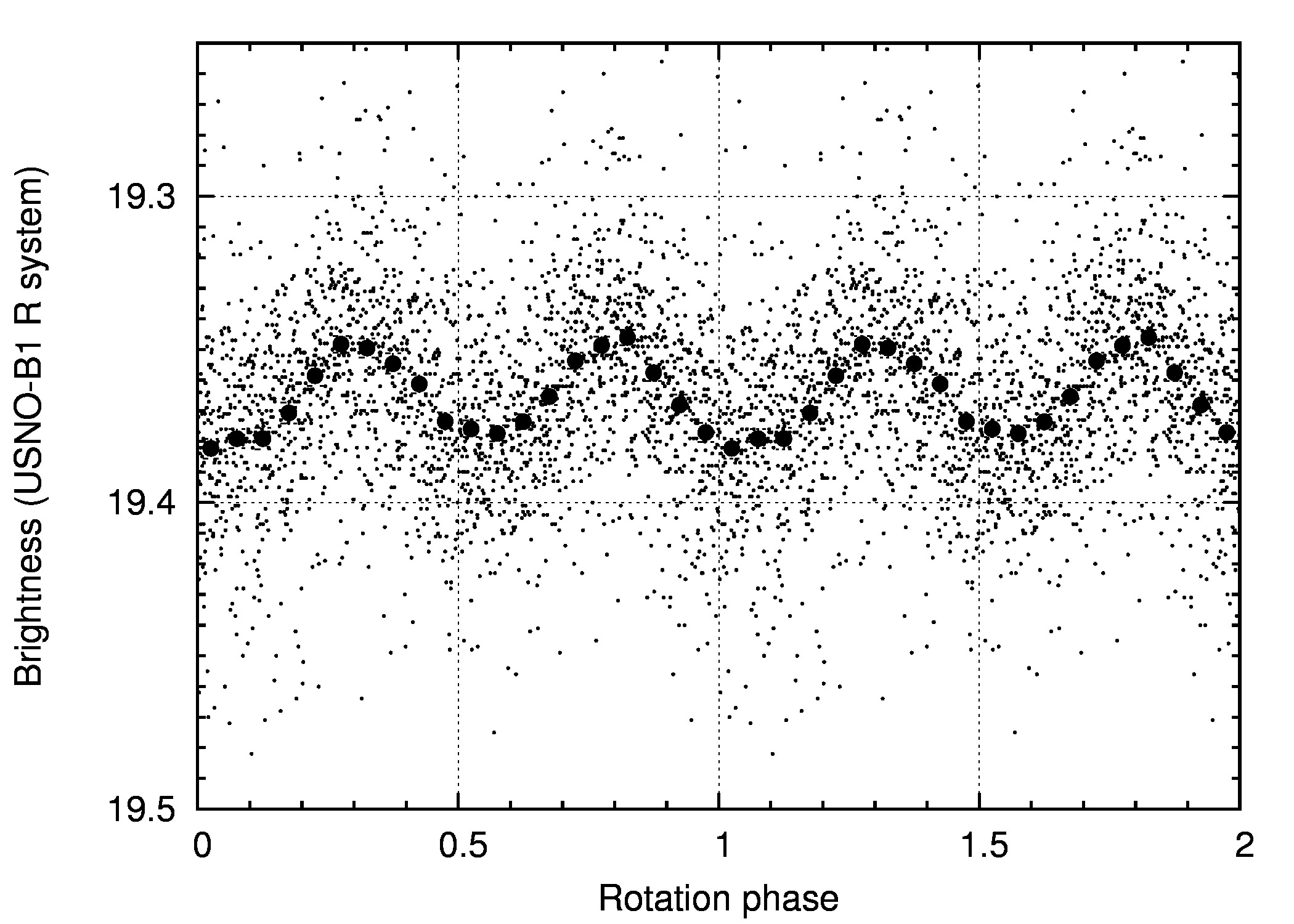}
\centering\includegraphics[width=0.45\textwidth]{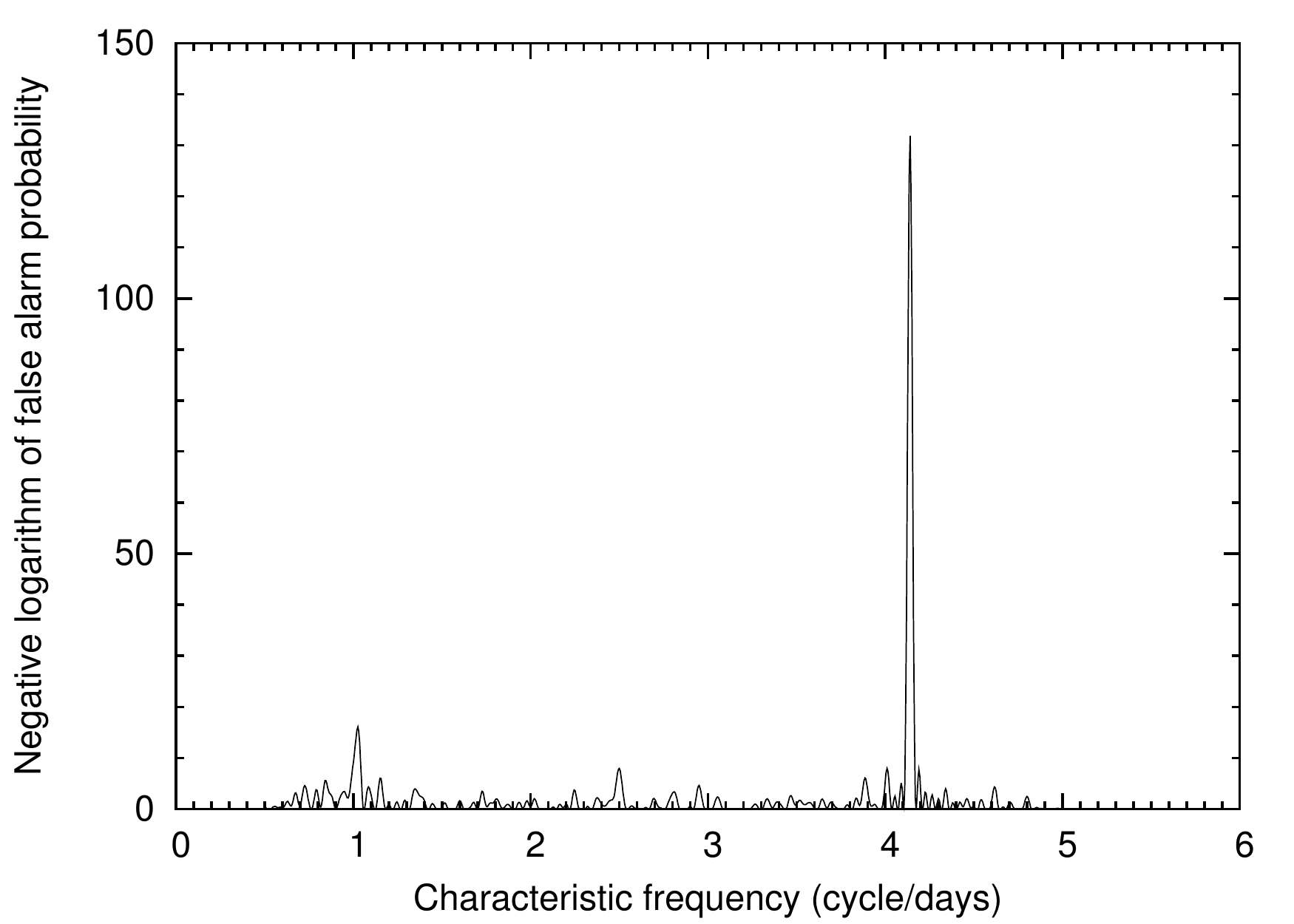}
\caption[]{Upper panel: Folded and binned light curve of Nereid as 
reconstructed from {\it Kepler}/K2 data after subtracting the 
long-term trend caused by the increasing phase angle and distance.
The binned light curve is shown above with $N=20$ bins, the formal uncertainties of the 
phase-binned data poins are in the range of 0\fm0027--0\fm0033.
The light curve shown in this plot is normalized to the brightness 
at $T=2,457,010\,{\rm JD}$, approximately 
at the center of the observations. The light curve is folded with a period of
$P=11.594\,{\rm h}$, corresponding to the double of the main frequency peak
at $n=4.140\,{\rm c/d}$ in the lower figure panel. 
Lower panel: Frequency spectrum computed from the {\it Kepler}/K2 light curve. 
The prominent peak at $n=4.140\pm0.006\,{\rm cycles/day}$ can be 
easily recognized.}
\label{fig:lc}
\end{figure}

Based on previous studies in the literature, 
the expected amplitude of the light curve variations are relatively small. 
Therefore, periodicity in the light curve has been searched by 
assuming a sinusoidal function coadded to a linear function representing
the gradual fading of the object (due to phase angle and distance variations).
This function was defined as 
\begin{equation}
A+B\Delta t+C\cos(2\pi n\Delta t)+D\sin(2\pi n\Delta t),
\end{equation}
where $n$ is the suspected rotational frequency and $\Delta t$ is the 
time after $2,457,010\,{\rm JD}$ (chosen in order to minimize 
the rounding errors and numerical artifacts). The unknown parameters
$A$, $B$, $C$ and $D$ can then be derived using linear regression. The
parameter space in $n$ is then scanned in the physically relevant domain
with a stepsize comparable to one tenth of the reciprocal timespan of the
observations (i.e. $0.001\,{\rm d}^{-1}$). 
A prominent peak has been detected
in the fit residuals at $n\,=\,4.140\pm0.006\,{\rm c/d}$,
as it can also be well seen in Fig.~\ref{fig:lc}b.
The corresponding values for the four coefficients at this peak was found to
be $\mathbf A=19.3650\pm0.0006$, $B=0.00335\pm0.00003$, $C=0.0129\pm0.0008$ and
$D=0.0101\pm0.0008$. The hypotenuse of $C$ and $D$, i.e. $\sqrt{C^2+D^2}$
gives us an estimation for the total amplitude which is then $0.0164\pm 0.0008$.
The residual of the fit was found to be $0.031$\,magnitudes.
Due to the significance of this peak,
we conclude that the rotation period is either $P=5.797\pm0.008\,{\rm h}$
or its double, $P=11.594\pm0.017\,{\rm h}$, assuming a double-peaked
solution. The amplitude of the light curve is 0\fm0164$\pm$0\fm0008
(0\fm0328$\pm$0\fm0018 peak-to-peak). To correct for phase angle effects, a 
phase constant of k\,=\,0.123$\pm$0.005\,mag\,deg$^{-1}$ was also derived
from our K2 data (Nereid was seen at phase angles of \lele{1\fdg2}{$\alpha$}{1\fdg9}
during the K2 observations).

The light curve we found is very similar to those
obtained by \cite{Grav2003} and \citet{Terai}, both in rotation period 
(11\fh52$\pm$0\fh14 and 11\fh50$\pm$0\fh10), and
amplitude (0\fm029$\pm$0\fm003 and 0\fm031$\pm$0\fm001).
In Sect.~\ref{sect:spin} we analyse in detail 
the consequences of the similarities of these light curves obtained
over a $\sim$15-year period. 

%%%%%
%%%%%
%%%%%%%%%%%%%%%%%%%%%%%%%%%%%%%%%%%%%%%%%%%%%%%%%%%%%%%%%%%%%%%%%%%%%%

\subsection{Spin axis constraints}
\label{sect:spin}

%%%%%
%\begin{figure}
%\centering\includegraphics[width=0.45\textwidth]{nereid_shape_2.png}
%\caption[]{A shape model of Nereid that matches the observed light curve, assuming 
%a subsolar latitude of 60\degr. The main axes of the body are marked (a\,$>$\,b\,$>$c), 
%and the direction of the Sun is indicated by the thick, black arrow ($\phi$\,=\,30\degr in
%the case of the present model))}
%\label{fig:lc}
%\end{figure}
%%%%%
%%%%%
\begin{figure}
\centering\includegraphics[width=8cm,angle=0]{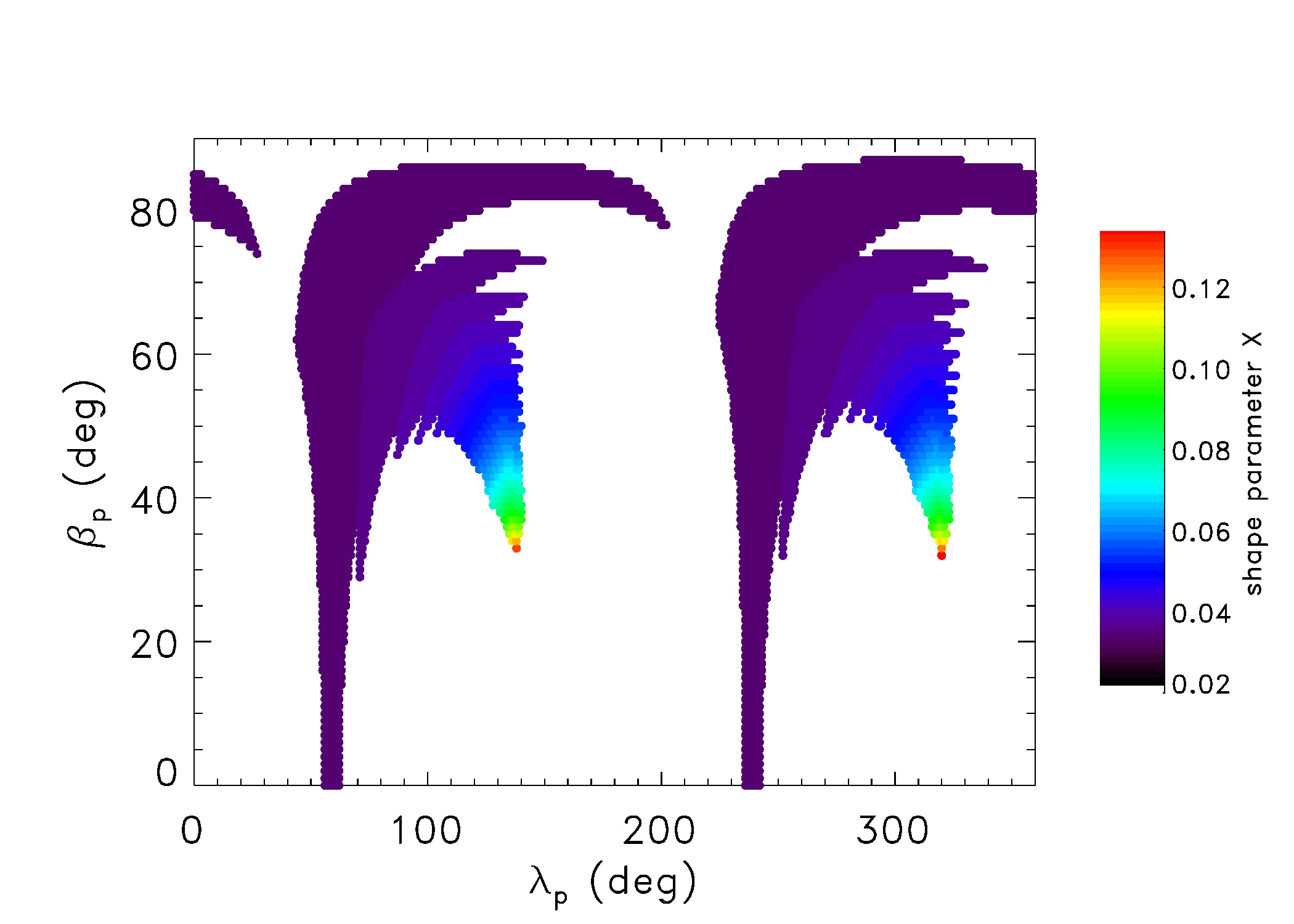}
\centering\includegraphics[width=8cm,angle=0]{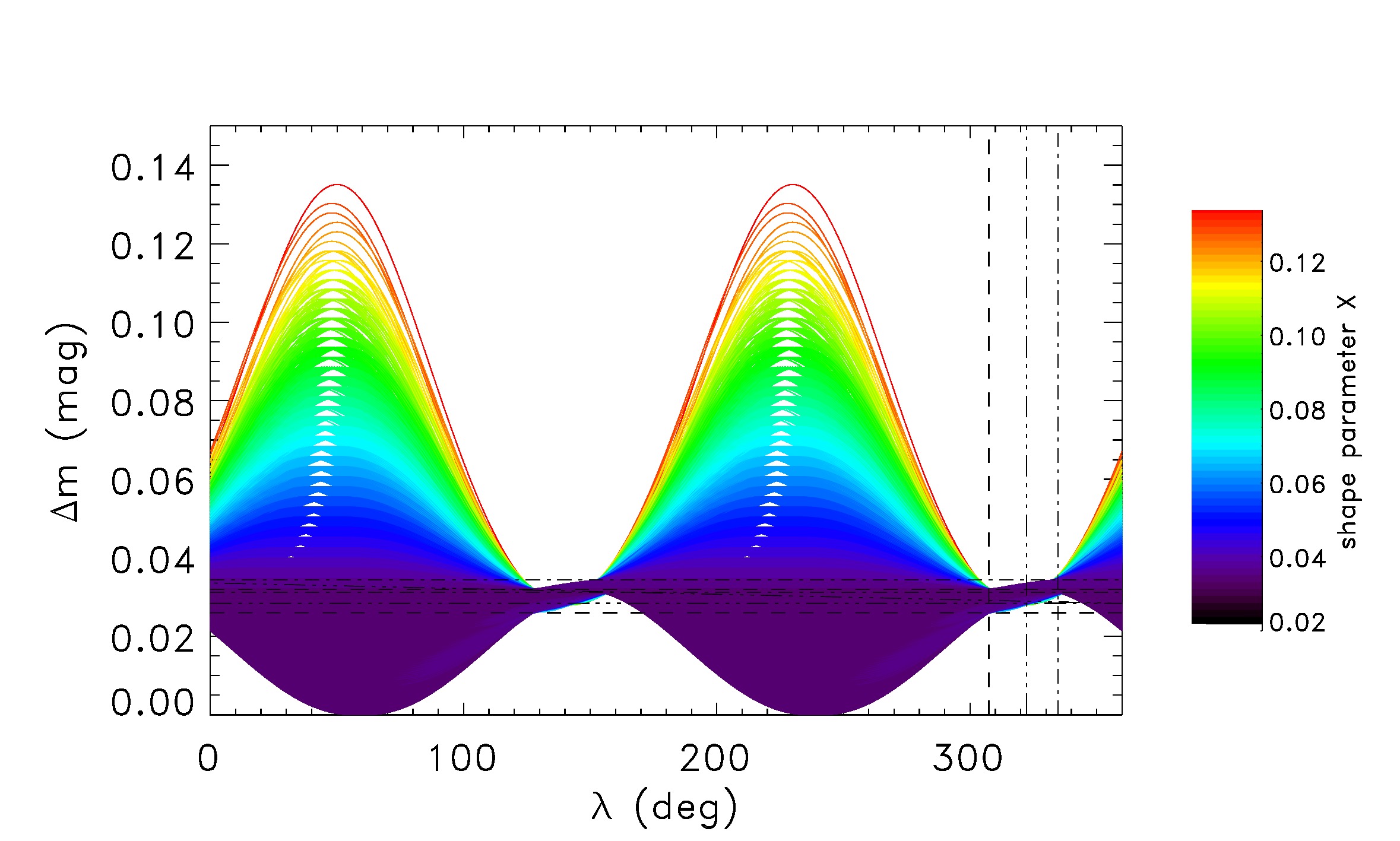}
\caption{Possible spin axis geometry configurations that reproduce the observed 
light curve amplitudes in 2001, 2008 and 2015. 
Upper panel: Allowed ecliptic longitude and latitude of the spin axis. The
colours correspond to the different X axis ratio parameters from the lowest (purple) 
to the highest (red, see the text for more details). 
Lower panel: Variation of light curve amplitude with the ecliptic longitude of 
Nereid assuming a stable spin axis orientation. The colours correspond to different 
shapes with X axis ratio parameters ranging from 0.03 to 0.14, as above. 
The dashed and dash-dotted \emph{vertical} lines mark the 
ecliptic longitude of the 2001, 2008 and 2015 observations, respectively. 
The dashed and dash-dotted \emph{horizontal} lines mark the rotation curve amplitude 
ranges allowed by the 2001, 2008 and 2015 observations, respectively. }
\label{fig:geoms}
\end{figure}
%%%%%%

The similarity of the 2001, 2008 and 2015 light curves puts constraints 
on the actual spin axis orientation. First, due to the short rotation period
with respect to the orbital period ($\sim$360\,days) the precession time 
of the spin axis is long, at least a few hundred years, even if very 
elongated body and/or high original obliquity are assumed \citep{Dobrovolskis}. 
This suggests that the spin axis of Nereid remained in approximately the
same direction in the last $\sim$15\,years. However, Neptune (and hence Nereid) 
moved on its orbit about 30\degr{} along the ecliptic in the last 15 years, 
and therefore the aspect angle $\vartheta$ of the spin axis 
should have changed with respect to 
the observer even if it was otherwise pointing in the same direction in space
in this period. The position angle $\vartheta$
depends on the ecliptic coordinates of the target ($\lambda$, $\beta$) as well
as on that of the spin axis' direction ($\lambda_p$, $\beta_p$): 
%%%%%%%%%%%%%%%
\begin{equation}
\cos\vartheta = -\sin \beta \, \sin \beta_p - \cos \beta \, \cos \beta_p \, \cos(\lambda-\lambda_p)
\label{eq:theta}
\end{equation}
%%%%%%%%%%%%%%%
\noindent If we assume that the shape of Nereid is a triaxial ellipsoid 
with semi-axes of $a$\,$>$\,$b$\,$>$\,$c$ then the object 
seen at a $\vartheta$ spin axis aspect angle will show a light curve 
with an amplitude of:
%%%%%%%%%%%%%%%
\begin{equation}
\Delta m = 2.5 log \sqrt{ {(b/c)^2\cos^2\vartheta + \sin^2\vartheta}\over{ 
(b/c)^2\cos^2\vartheta}+(b/a)^2\sin^2\vartheta }
\label{eq:theta}
\end{equation}
%%%%%%%%%%%%%%%   
\noindent Here we assumed that the light curve is \emph{solely caused by 
shape effects} and there are no albedo variegations on the surface.  
Due to the small phase angles the phase angle correction of the light curve amplitude 
is negligible. We characterise the semi-axis ratios of the triaxial ellipsiod 
by an axis ratio parameter $X$. With this parameter the semi-axes of the ellipsoid are:
\begin{equation}
b = 1,~~a = (1+X)b,~~c = (1-X)b 
\end{equation}
\noindent and the body rotates around its shortest ($c$) axis. 

With this assumption, and using the equations above, 
we calculated those \{$\lambda_p,\beta_p$\} combinations for which
$\Delta m$ corresponds to the observed light curve amplitues by 
\citet{Grav2003} in 2001-2002 ($\Delta m$\,=\,0\fm029$\pm$0\fm003),
\citet{Terai} in 2008 ($\Delta m$\,=\,0\fm031$\pm$0\fm001)
and by us in 2015 ($\Delta m$\,=\,0\fm0328$\pm$0\fm0018)
within the given uncertainties. In the case of the \citet{Terai} data 
we found the originally quoted uncertainties to be too optimistic. 
Our analysis of their photometry gives an amplitude uncertainty of 0\fm0026. 
We used this latter value in the spin axis orientation analysis. 

We originally assumed X in the 0.03 -- 0.34 range.
The lowest value of X corresponds to the smallest possible X value that could
produce a light curve compatible with the observations if $\vartheta$\,=\,90\degr{},
while the highest value corresponds to a $\sim$2:1 $a$:$c$ axis ratio,
suggested e.g. by \citet{Hesselbrock2013}. 
The results are plotted in Fig.~\ref{fig:geoms}
where the different colours mark the different axis ratio parameter values. 
Most of the allowed \{$\lambda_p,\beta_p$\} pairs belong to a small 
X value, i.e. to a slightly elongated shape with an axis ratio of 
$a$:$c$\,$\approx$\,1.06:1. The largest possible axis ratio parameter 
value we obtained is X\,=\,0.133. Larger X values are not allowed
by the amplitude constraints and hence more elongated shapes
(X\,$>$0.133) are excluded. 

We checked how the light curve amplitude evolves in the different cases allowed
by the 2001, 2008 and 2015 light curve amplitude constraints. In  Fig.~\ref{fig:geoms}b 
we present the light curve amplitude variation as a function of ecliptic longitude. 
At the observation epochs between 2001 and 2015 (ecliptic coordinates marked by vertical lines 
in the figure) the light curve amplitude does not change notably and remains low, irrespective 
of the shape of Nereid. While there are many solutions in which the amplitude remains low 
(slightly non-spherical cases, purple curves) along the solar orbit of the
Neptune-Nereid system, there are 
also cases when Nereid appears notably elongated and the light curve amplitude 
changes remarkably with the orbital phase around the Sun (cases marked with green to red). 
In these latter cases Nereid could have had its highest amplitude period 
in the 1960's, and still could have notably higher amplitudes in the 1980's than 
what we can see today. Another peak in the light curve amplitude would be expected in 
$\sim$30 years. The maximum possible light curve amplitude is 
$\Delta m$\,=\,0\fm13. 

%%%%%
\begin{figure}
\centering\includegraphics[width=8cm,angle=0]{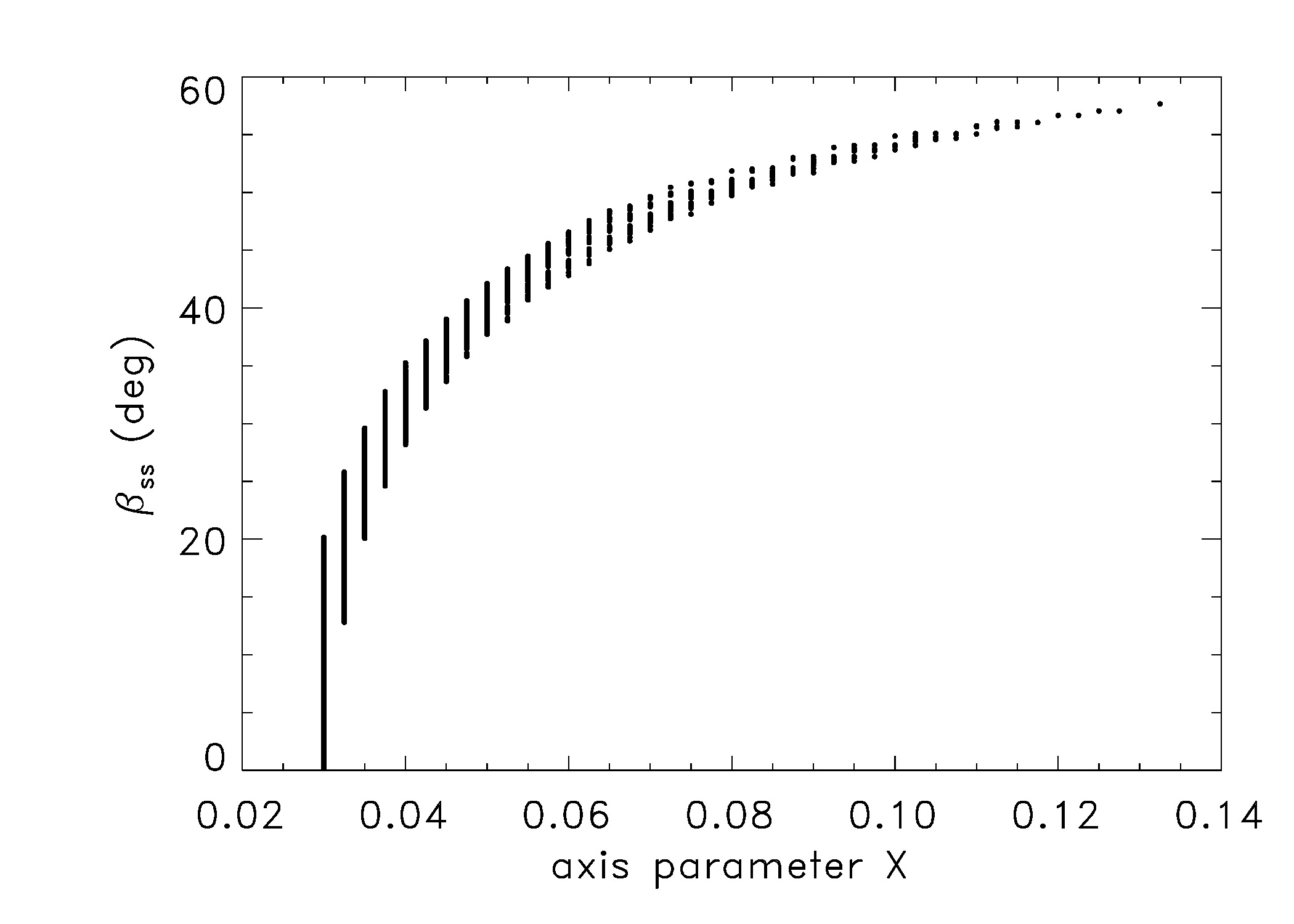}
\centering\includegraphics[width=8cm,angle=0]{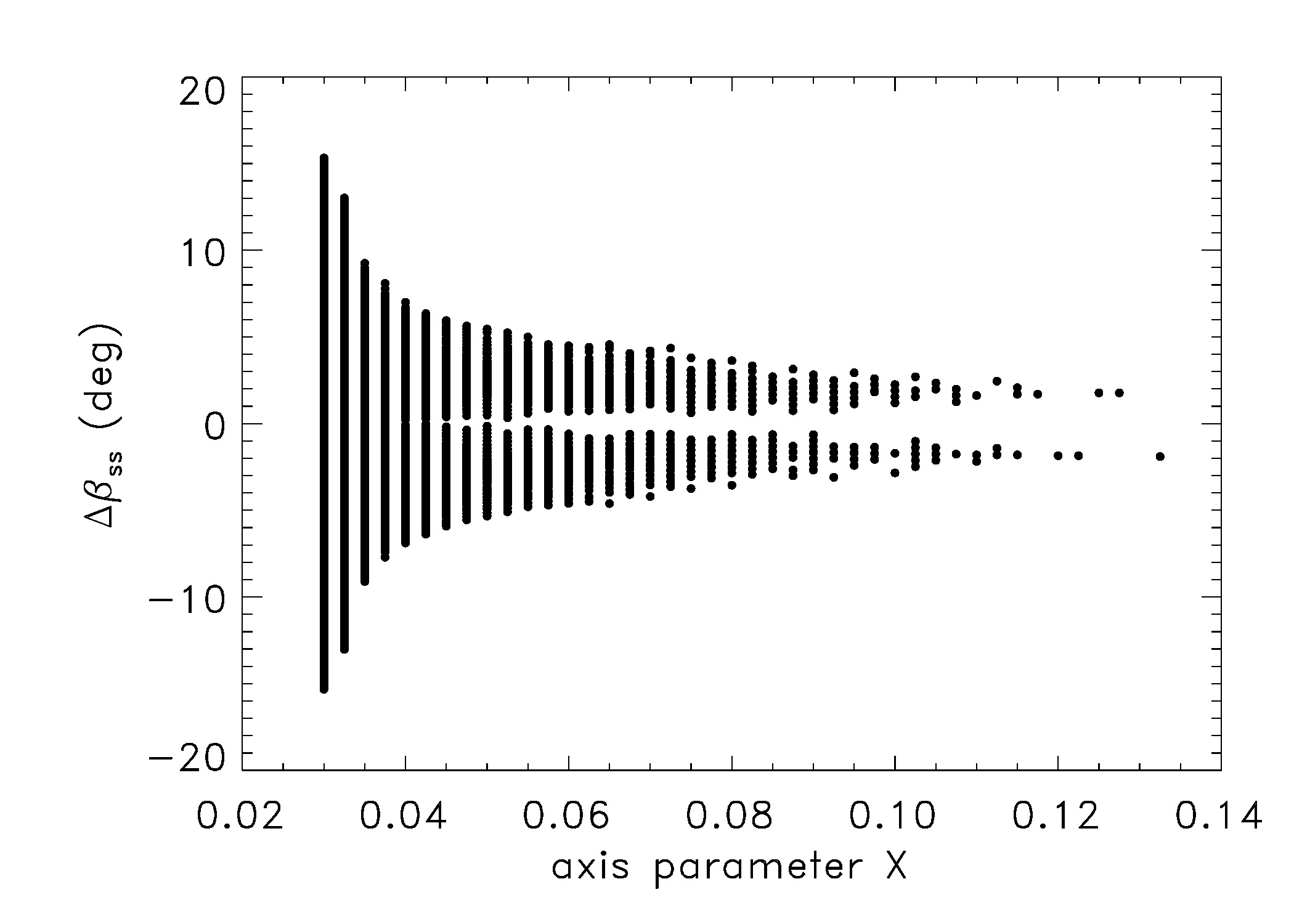}
\caption{Upper panel: Subsolar latitude ($\beta_{ss}$) for the allowed 
geometry and shape configurations as a function of the axis ratio parameter X, 
at the Spitzer/MIPS epoch. 
Lower panel: Variation of the subsolar latitude between the 
Spitzer/MIPS and the Herschel/PACS observation epochs as a function of the 
axis ratio parameter X. }
\label{fig:vartheta}
\end{figure}
%%%%%%

For the thermal emission models that we discuss in Sects.~\ref{sect:neatm}
and \ref{sect:tpm}
it is important to see how the spin axis aspect angles and the 
subsolar latitudes could change with time. We plotted the 
subsolar latitude $\beta_{ss}$ as a function of 
the axis ratio parameter X in Fig.~\ref{fig:vartheta}a for the allowed 
configurations at the epoch of the Spitzer/MIPS observations. 
The subsolar latitudes show a well defined relationship
with the axis ratio parameter: at low X values $\beta_{ss}$ is close to 
zero (equator-on configurations) then it rises quickly and reaches a 
maximum value of $\beta_{ss}$\,$\approx$\,60\degr{}
for the largest possible values of X. 
While $\beta_{ss}$ can change notably between the MIPS and PACS epochs 
(up to $\Delta\beta_{ss}$\,$\approx$\,15\,\degr) for low
X values, the change is rather small for higher axis ratio parameters 
(see Fig.~\ref{fig:vartheta}b).

%%%%%%%%%%%%%%%%%%%%%%%%%%%%%%%%%%%%%%

\section{The thermal emission of Nereid}

\subsection{Herschel Space Observatory measurements \label{sect:Herschel}}

\begin{figure*}
\centering\includegraphics[width=\textwidth,angle=0]{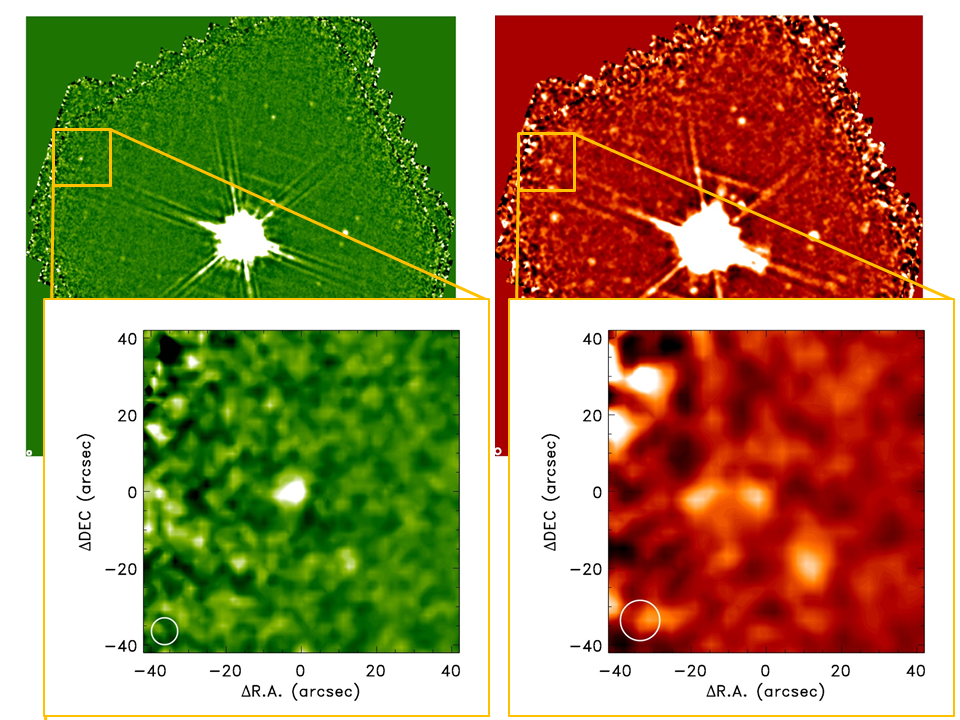}
\caption{Nereid on the Herschel/PACS 100\,$\mu$m (left) and 160\,$\mu$m, 
co-added images, reduced in the co-moving frame of Neptune.   
The large images show the full area mapped with Neptune in the centre, 
while the zoomings show the close environment of Nereid, with 
the moon in the centre of the stamps. At 160\,$\mu$m Nereid was identified 
by positional matching with other wavelengths from 
the multiple sources of similar brightness.}
\label{fig:herschelimage}
\end{figure*}

We have found Nereid in archival Herschel Space Observations data 
(proposal ID: OT1\_ddan01\_1). 
The original target in these observations was Neptune, and Nereid was just 
accidentally on the images (see Fig.~\ref{fig:herschelimage}). Four
observations were taken with the PACS 
\citep[Photometer Array Camera and Spectrometer][]{pacs} photometer camera
on board the Herschel Space Observatory \citep{herschel},
using the 100/160\,$\mu$m filter combination in all four cases.  
The data reduction pipeline we used is the same as 
the one used in the "TNOs are Cool!" Herschel Open Time Key Program
\citep{Mueller2009}, described in detail in \citet{Kiss2014}. As our aim
was to obtain photometry of a point source, we used
the photProject() task with high pass filtering to create maps from 
the time domain detector data. The images of the four consecutive measurements
were stacked in the co-moving frame of Nereid, and aperture photometry 
was performed on the stacked 100 and 160\,$\mu$m images. 
Flux uncertainties were determined with the implanted source method
\citep{Kiss2014}, but in this case using a $\sim$80\arcsec area 
around Nereid rather than the high coverage regions of the whole map.
As the apparent movement of the target was very small during the measurements 
we were not able the perform any kind of background correction. 
We clearly detected Nereid in both bands, and the photometry provided
F$_{100}$\,=22.8$\pm$1.7\,mJy and F$_{160}$\,=18.6\,$\pm$2.9\,mJy 
in-band flux densities at 100 and 160\,$\mu$m, respectively. 

%%%%%%%%%%%%%%%%%%%%%%%%%%%%%%%%%%%%%
\begin{table*}
\caption{Summary of Nereid's thermal infrared observations. 
The "identifier" column refers to OBSID in the case of PACS and AORKEY
in the case of MIPS measurements. }
\label{table:obs}
\begin{tabular}{clcrrccc}
Instrument & Obs. date & identifier & duration & filter combination & $r_h$ & $\Delta$ & $\alpha$ \\
           &  (JD)     &            & (s)      &  ($\mu$m/$\mu$m)   &  (au) & (au)     &  (deg) \\
\hline
Herschel/PACS & 2456090.655 & 1342222561 & 2996 & 100/160 & 30.02 & 29.65 & 1.85 \\
Herschel/PACS & 2456090.691 & 1342222562 & 2996 & 100/160 &       &       & \\
Herschel/PACS & 2456090.726 & 1342222563 & 2996 & 100/160 &       &       & \\
Herschel/PACS & 2456090.679 & 1342222564 & 2996 & 100/160 &       &       & \\
\hline
Spitzer/MIPS  & 2453539.667 & 4535808    & 1012 & 24 & 30.07 & 29.57 & 1.70 \\
Spitzer/MIPS  & 2453539.679 & 4535808    & 1012 & 70 &       &       & \\
%\hline
%\tablecomments{Table \ref{table:phot} is published in its entirety in the
%electronic edition of the {\it Astrophysical Journal Letters}.  A portion is
%shown here for guidance regarding its form and content.}
%\tablenotetext{a}{Magnitudes shown here are transformed to USNO-B1.0 $R$
%system, see text for further details.}
\hline
\end{tabular}
\end{table*}
%%%%%%%%%%%%%%%%%%%%%%%%%%%%%%%%%%%%%%%%

%%%%%
%%%%%%

\subsection{Spitzer/MIPS measurements \label{sect:Spitzer}}

\begin{figure*}
\centering{\hbox{
\includegraphics[width=5.5cm,angle=0]{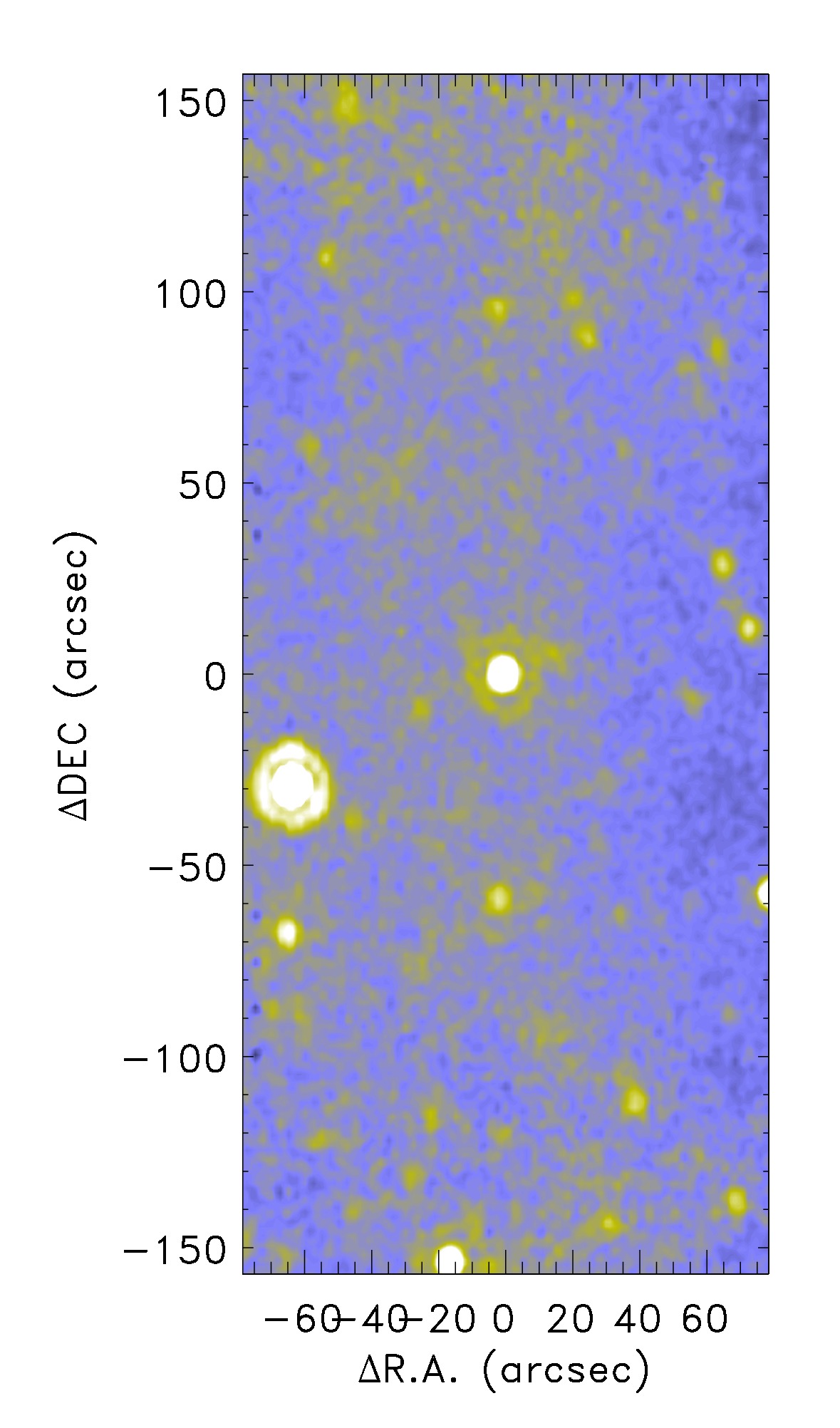}
\includegraphics[width=5.5cm,angle=0]{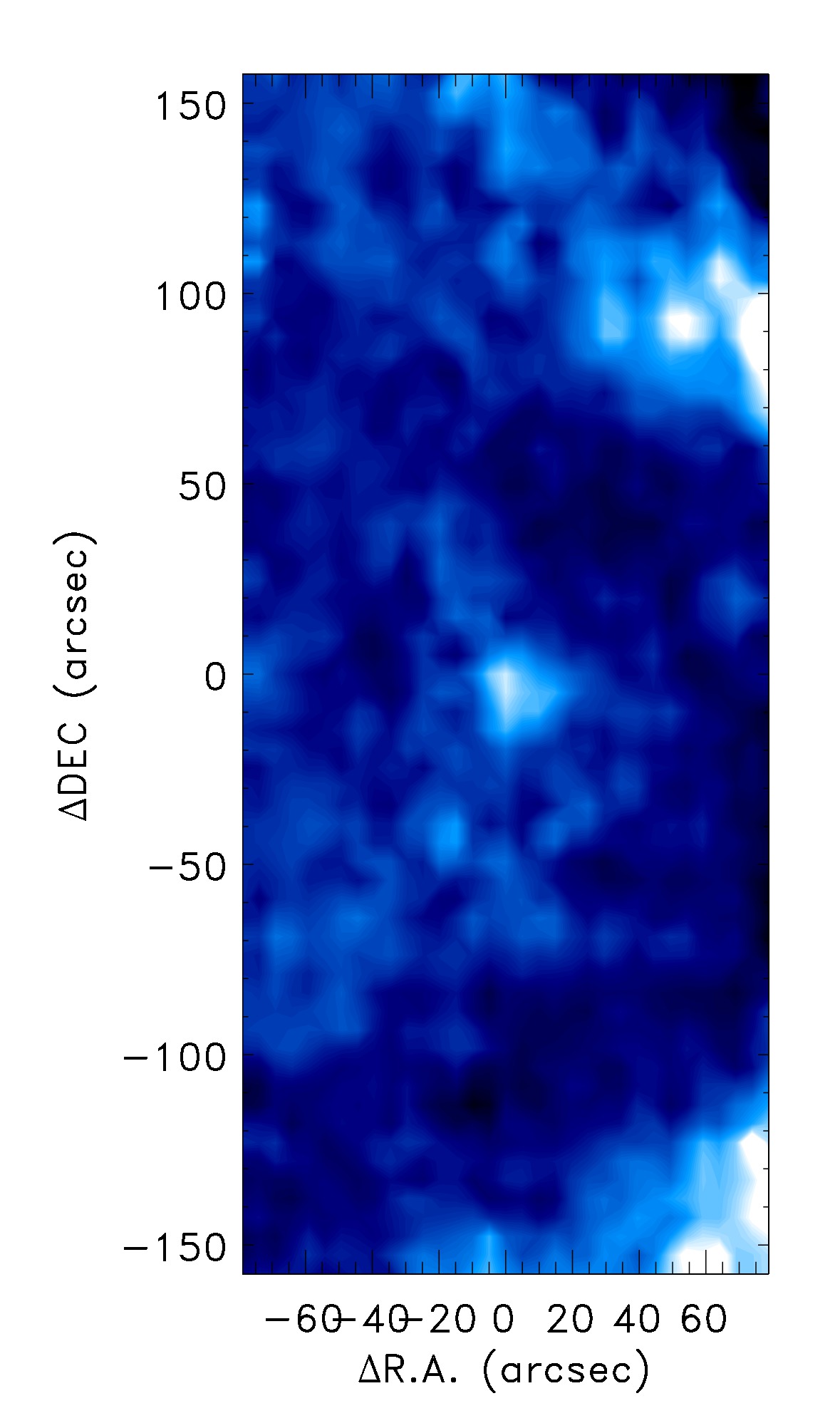}
\includegraphics[width=5.5cm,angle=0]{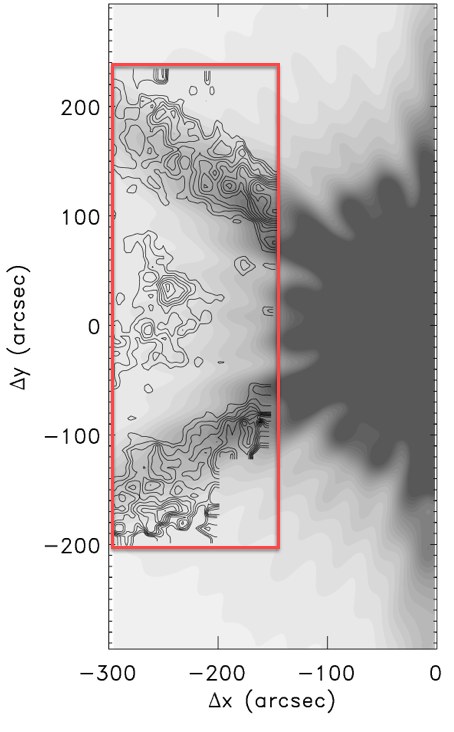}
}}
\caption{Nereid as observed at  24 and 70\,$\mu$m with the MIPS instrument
of the Spitzer Space Telescope. Left panel: 24\,$\mu$m false colour image.
Middle panel: 70\,$\mu$m false colour image; Nereid is the bright compact source
in the centre of both images. Right panel: MIPS 70\,$\mu$m intensity 
contours (the same as the middle panel) 
overlaid on an intensity map (grayscale) of a model image of Neptune which
was not in the field-of-view. The red rectangle shows the area of 
the MIPS 70\,$\mu$m image. 
The position of the bright source relative to Nereid is set in such a way that it 
corresponds to the relative position of Nereid and Neptune at the time 
of the Spitzer/MIPS observations. }
\label{fig:mipsimage}
\end{figure*}

The Spitzer Space Telescope observed Nereid in a dedicated observation
in 2005 using the MIPS camera at 24 and 70\,$\mu$m \citep{MIPS}. 
The data were reduced using the same pipeline as was used for the reduction of 
MIPS data of Centaurs and trans-Neptunian objects by 
\citet{Migo} and \citet{stansberry2008,stansberry2012}. 
Nereid was clearly detected in both bands, and we obtained in-band flux densities of 
F$_{24}$\,=\,2.56$\pm$0.03\,mJy at 24\,$\mu$m and F$_{70}$\,=\,50$\pm$11\,mJy at 
70\,$\mu$m using multi-aperture photometry. The corresponding images are presented 
in Fig.~\ref{fig:mipsimage}. At the time of the observations Nereid was at  
$\sim$4\arcmin{} separation from the very bright
Neptune. While the 24\,$\mu$m image (Fig.~\ref{fig:mipsimage}a) does not seem to be 
contaminated by extended emission features from the planet, this is not the case at 
70\,$\mu$m. The two bright rims at the
top and bottom of the 70\,$\mu$m image (Fig.~\ref{fig:mipsimage}b) are caused by the 
hexagon structure of the Neptunian point spread function (PSF), 
as it is demonstrated in Fig.~\ref{fig:mipsimage}c. 
There are also less apparent emission spikes at the location of Nereid that may affect the
photometry. To account for the emission from Neptune at the location of Nereid, 
we tried to subtract Neptune's contribution by scaling a theoretical 70\,$\mu$m 75\,K PSF,
scaled to the actual brightness of Neptune in this band at the time of the observations. 
Neptune's brightness was estimated using the Neptune model by R.~Moreno
\citep{Moreno} and we obtained F$_{70}^{Neptune}$\,=\,383.2\,Jy at 
the effective wavelength of 71.42\,$\mu$m of this MIPS band.
This model has been used for the Herschel/PACS and SPIRE flux calibration and used to
reproduce the flux densities of Neptune within 2\% (M\"uller et al., A\&A, submitted). 
With this correction the main emission features have disappeared and repeated photometry 
provided an in-band flux of F$'_{70}$\,=\,29$\pm$16\,mJy. 
  
%%%%%%
%%%%%%

%%%%%
\begin{table}
\begin{tabular}{ccccc}
\hline
Detector/  & $\lambda_{eff}$ & F$_i$  & C$_\lambda$ &  F$_m$  \\ 
filter     &   ($\mu$)m      & (mJy)  &            &    (mJy) \\        
\hline
MIPS 24   &  23.68   & 2.56$\pm$0.03 &  0.99$\pm$0.01 &   2.59$\pm$0.13  \\     
MIPS 70/u &  71.42   & 50$\pm$11     &  0.91$\pm$0.01 &   55.0$\pm$12.7  \\
MIPS 70/c &  71.42   & 29$\pm$16     &  0.91$\pm$0.01 &   31.9$\pm$17.6  \\  
PACS 100  &  100.0   & 22.8$\pm$1.7  &  0.99$\pm$0.01 &   23.0$\pm$2.1 \\  
PACS 160  &  160.0   & 18.6$\pm$2.9  &  1.03$\pm$0.01 &   18.1$\pm$3.0 \\  
\hline
\end{tabular}
\caption[]{Spitzer/MIPS and Herschel/PACS photometry results. 
The columns of the table are: 
Detector and filter combination; 
F$_i$: inband flux, as determined from the corresponding image;
$\lambda_{eff}$: effective wavelength of the band;
C$_\lambda$: Colour correction factor; 
F$_m$: final, monochromatic flux, used for radiometry modelling. The
F$_m$ monochromatic fluxes are obtained as 
F$_m$($\lambda$)\,=\,F$_m$($\lambda$)/C$_\lambda$. 
MIPS 70/u and 70/c refer to the uncorrected and corrected 70\,$\mu$m fluxes, 
as described in the text. An instrumental calibration accuracy of 
5\% was assumed for all detector/filter combinations and added 
quadratically to the flux uncertainties.}
\label{table:fluxes}
\end{table}
%%%%%

%\section{Radiometry}
%\label{sect:radiometry}

\subsection{NEATM model}
\label{sect:neatm}

%%%%%%%%%%%%%%%%%%%%%
\begin{figure}
\centering\includegraphics[width=8cm,angle=0]{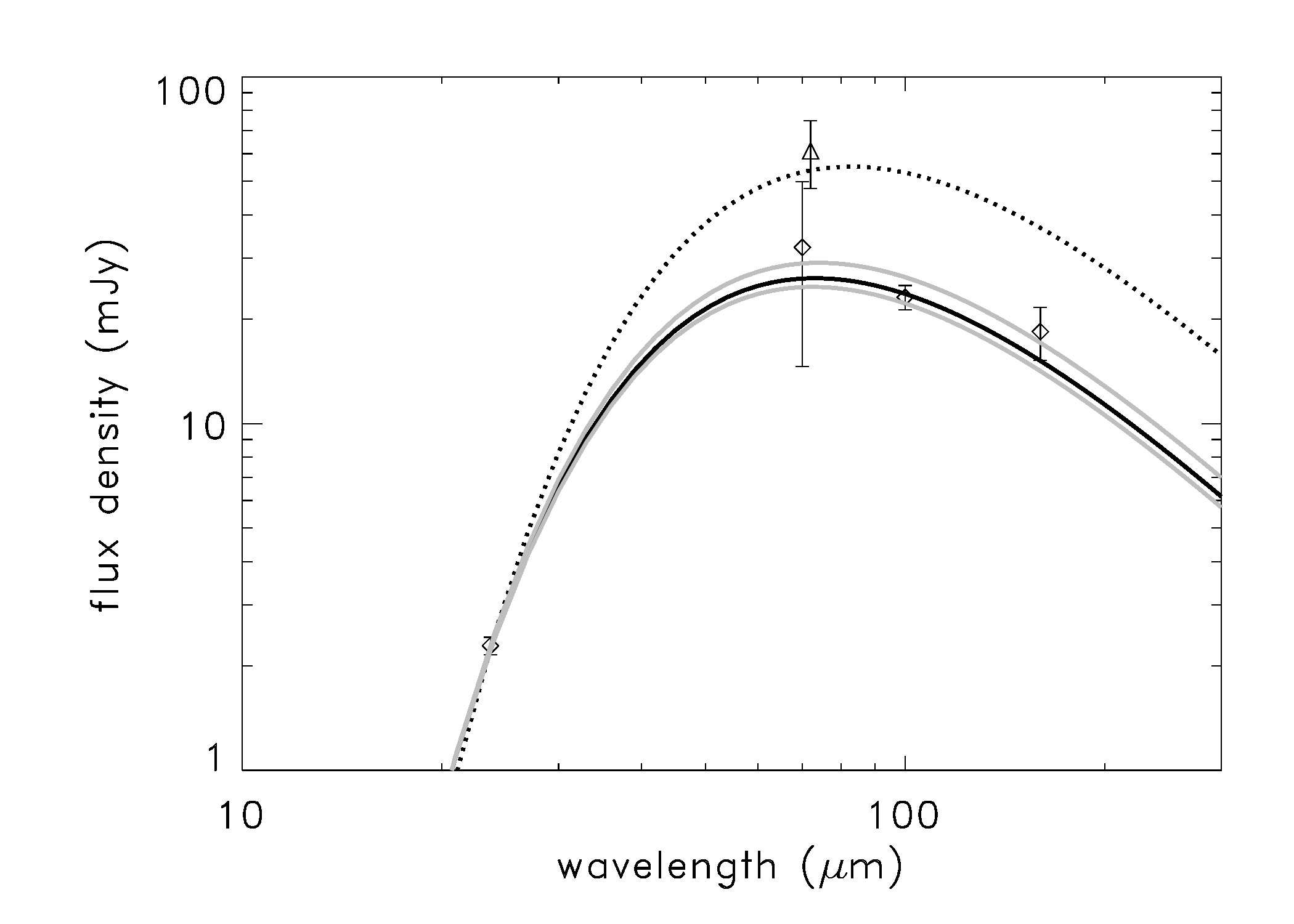}
\centering\includegraphics[width=8cm,angle=0]{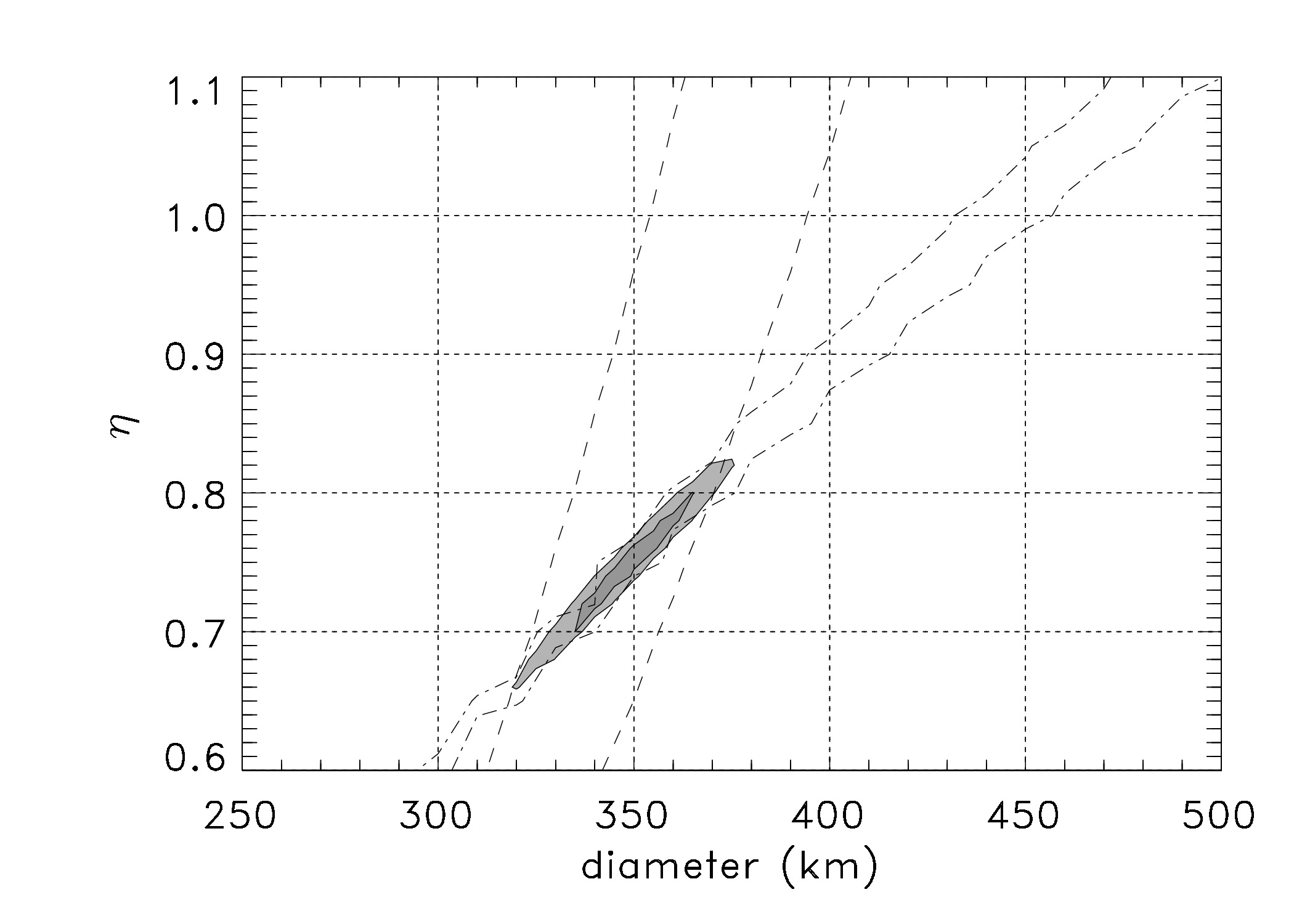}
\caption{Upper panel: Best-fit NEATM model of the thermal emission of
Nereid with the observed and colour corrected 
fluxes overplotted (Case $d$, diamond symbols and solid line). 
The $\chi^2$\,=\,(1+$\sigma$)$^2$ NEATM fits are drawn by gray curves. 
The NEATM fit using the uncorrected MIPS 70\,$\mu$m flux, corresponding to
Case $a$, is also plotted with a dotted curve. The uncorrected 70\,$\mu$m  
plot is presented by a triangle and slightly shifted in wavelength from 
the uncorrected point for clarity.  
Lower panel: Reduced $\chi^2$ contour map of the NEATM model fits as a function
of the effective diameter and the beaming parameter $\eta$. The outermost 
contour of 1.7 correspond to the reduced $\chi^2$ limit of acceptable 
models with four data points and two model parameters (diameter and $\eta$)}
\label{fig:neatmres}
\end{figure}
%%%%%%%%%%%%%%%%%%%%%

We used the Near-Earth Asteroid Thermal Model (NEATM) 
\citep{harris1998} to estimate the thermal emission of 
Nereid using the Spitzer/MIPS 24 and 70\,$\mu$m and the Herschel/PACS 100 and 160\,$\mu$m 
fluxes, as listed in Table~\ref{table:fluxes}. The Spitzer/MIPS and Herschel/PACS
observations are separated by about six years, and, as we have seen above, there were
indications that the rotational axis, brightness and the apparent shape 
of Nereid may change significantly on this timescale (see Sect.~1). 
However, as discussed above, the similarity of our newly detected rotation solution to
those by \citet{Grav2003} and \citet{Terai} indicates that Nereid has been in a low 
rotational light curve amplitude period in the recent decade that is also 
associated with relatively small changes in spin axis rotation angle and 
subsolar latitude configurations. Therefore we feel that the combination of the 
Spitzer/MIPS and Herschel/PACS data is likely feasible and can
better constrain the radiometry solutions than analysing them separately. 

We characterise the radiometry solutions by calculating the reduced $\chi^2$ values 
of the fits from the observed and model fluxes of a specific effective diameter
and beaming parameter combination. We consider a solution 
acceptable if $\chi^2$\,$\le$\,$(1+\sigma)^2$, where $\sigma$ is the
standard deviation of the $\chi^2$ distribution
\citep[see e.g.][for a similar application of the method]{Vilenius2014}. 
Albedo and effective diameter are not independent
but are linked by the absolute magnitude. We use H$_V$\,=\,4\fm418$\pm$0\fm008, 
a weighted average of the absolute magnitudes provided by \citet{Grav2004} and
\citet{Rabinowitz2007}. 

For the sake of completeness, we derived four kind of fits:
%%%
\begin{itemize}
%%%
\item[a)] \emph{MIPS 24 and 70\,$\mu$m only, original 70\,$\mu$m flux (MIPS epoch)}. 
In this case we used input flux derived from the uncorrected 
70\,$\mu$m flux, F$_m(70)$\,=\,55.0$\pm$12.7\,mJy (see Table~\ref{table:fluxes}). 
The NEATM fits resulted in a best fit diameter of 
D\,=\,605$\pm$95\,km, $\eta$\,=\,1.36$\pm$0.16, and p$_V$\,=\,0.081$\pm$0.027.
(see the corresponding SED, plotted with dotted line, in Fig.~\ref{fig:neatmres}a).
This diameter is by far larger than any conceivable size for Nereid, and clearly
indicates that the MIPS 70$\mu$m flux is strongly contaminated, 
most likely by the diffraction spikes of Neptune, as discussed in Sect.~\ref{sect:Spitzer}
earlier. Therefore this 70\,$\mu$m input flux and the related solution is not considered 
any further in our analysis. 
%%%
\item[b)] \emph{MIPS 24 and 70\,$\mu$m only, corrected 70\,$\mu$m flux (MIPS epoch)}.  
In Fig.~\ref{fig:neatmres}b we plot the regions of the acceptable solution
($\chi^2$\,$\le$\,$(1+\sigma)^2$) in the diameter -- beaming parameter space. 
For Case $b$ this is represented by the area enclosed by dashed lines. 
Due to the large error bar of the 70\,$\mu$m flux a wide range of diameters,
and an especially wide range of beaming parameters are allowed. 
%
%The best fit NEATM solution provides D\,=\,405$\pm$95\,km, 
%$\eta$\,=\,0.90$\pm$0.20, and p$_V$\,=\,0.18$\pm$. Although this solution 
%is compatible with e.g. the Voyager-2 flyby diameter (350\,km), there is
%a considerable uncertainty in the solutions, driven by the large error of the 70\,$\mu$m flux. 
%%%
\item[c)] \emph{PACS 100 and 160\,$\mu$m only (PACS epoch)}. When only the PACS 
measurements are considered the situation is somewhat different. 
Here we have a strong correlation of the diameter and
$\eta$, but the model is not well constrained due to the lack of short 
wavelength fluxes (region enclosed by dash-dotted lines in 
Fig.~\ref{fig:neatmres}{\bf{b}}). 
%
%D\,=\,410$\pm$80\,km, 
%$\eta$\,=\,1.36$\pm$0.58, and p$_V$\,=\,0.18$\pm$. In this case we lack 
%short wavelength information hence the beaming parameter is not well constrained,
%which is reflected in the large $eta$ uncertainties. 
%%%
\item[d)] \emph{PACS and MIPS combined (corrected 70\,$\mu$m flux, 
both MIPS and PACS epochs).}
Due to the similarities of the observing geometries at the dates 
of the MIPS and PACS observations the resulting fluxes of the same 
NEATM model are almost equivalent for the two epochs, and therefore a mean 
observing geometry can be safely applied on the combined dataset
(see Table~\ref{table:obs}). 
When all four data points are considered, the NEATM model becomes well 
constrained. The common solution
(gray contour ellipses in Fig.~\ref{fig:neatmres}b) is compatible with
both the MIPS-only and the PACS-only solutions (Cases $b$ and $c$).  
In this case we obtain D\,=\,345$\pm$15\,km, 
$\eta$\,=\,0.76$\pm$0.02, and p$_V$\,=\,0.25$\pm$0.02. This effective diameter 
is very similar to the Voyager-2 flyby value of 350$\pm$50\,km \citep{Thomas1991}.
The low $\chi^2$ value obtained for this combined solution confirms that
the $\eta$ values related to the two epochs are the same to a level
that they cannot be distinguished with the current flux uncertainties.
\end{itemize}

We consider the combined dataset and the solution of Case $d$ 
as the most acceptable size, albedo and beaming parameter for Nereid.
It is supported by the light curve analysis / spin axis constraints and 
none of the other NEATM solutions (Cases $b$ and $c$) 
contradict with solution $d$ considering all
the errors in the NEATM model parameter determination. 

Using our favoured radiometry solution of D\,=\,345$\pm$15\,km, 
$\eta$\,=\,0.76$\pm$0.02, and following the relations of thermal parameter, beaming parameter and
surface roughness described in Spencer et al. (1989) and Spencer (1990),
we can constrain the thermal properties (thermal inertia and surface 
roughness) of Nereid using the NEATM solutions to some level.  
%based on the best fit beaming parameter $\eta^*$\,=\,0.76. 
In the framework of the abovementioned models, a 
beaming parameter below unity can be explained by surface roughness
effects. The minimum level of surface roughness beaming contribution
required to obtain $\eta$\,=\,0.76 (our best-fit value) is $\delta\eta$\,=\,0.24
(see eq. 7 in Spencer 1990) that corresponds to an r.m.s. surface 
roughness level of $\rho$\,=\,0.6 (Lagerros 1998). Such a surface
can also be constructed by considering 90\degr{} hemispheral 
craters with 50 per cent surface coverage. To obtain this low $\eta$, it is, however, 
required in addition to the moderately high surface roughness that 
either the thermal inertia is extremely low ($\Gamma$\,$<$\,0.1\,\tiunit) \emph{or} the
subsolar latitude is $\beta_{ss}$\,$\approx$\,90\degr{} (we see the spin-axis near to pole-on). 
For a near to pole-on configuration the thermal intertia should still be low, but it can be a value 
somewhat higher than in general case, typically $\Gamma$\,$\approx$\,2\,\tiunit{} may be possible. 
The other possibility, $\beta_{ss}$\,$\approx$\,90\degr{} is not supported by the light 
curve constraints, as a maximum value of $\beta_{ss}$\,$\approx$\,60\degr{}
is obtained in the spin axis analysis 
(see Sect.~\ref{sect:spin} and Fig.~\ref{fig:vartheta}). 

If we allow for a high surface roughness 
($\rho$\,=\,0.9 that can be achieved with 90\degr{} craters at 100\% coverage), 
different sub-solar latitudes become possible. E.g. for $\beta_{ss}$\,=\,0\degr{}
$\eta$\,=\,0.76 results in a very low thermal inertia of $\Gamma$\,=\,0.5\,\tiunit.
\citet{Lellouch2013} found an average thermal inertia of 
$\Gamma$\,=\,2.5$\pm$0.5\,\tiunit among Centaurs and trans-Neptunian objects
in the heliocentric distance range of 25\,au\,$<$\,r$_h$\,$<$\,41\,au, 
with a strong suggestion of decreasing $\Gamma$ with increasing heliocentric distance.
Even higher values of $\Gamma$ are expected among the icy moons of the outer giant
planets. This may indicate that the observed
values of the beaming parameter may rather be explained by a higher
subsolar latitude ($\beta_{ss}$\,$\approx$\,60\degr) than by a very low thermal 
inertia value of the surface. This also suggests that the shape of Nereid
may be elongated to some level, because high $\beta_{ss}$ values are found
for larger axis ratio parameters in Sect.~\ref{sect:spin}. 

\subsection{Thermophysical modelling}
\label{sect:tpm}

%%%%%%%%%%%%%%%%%%%%%
\begin{figure}
\centering\includegraphics[width=8cm,angle=0]{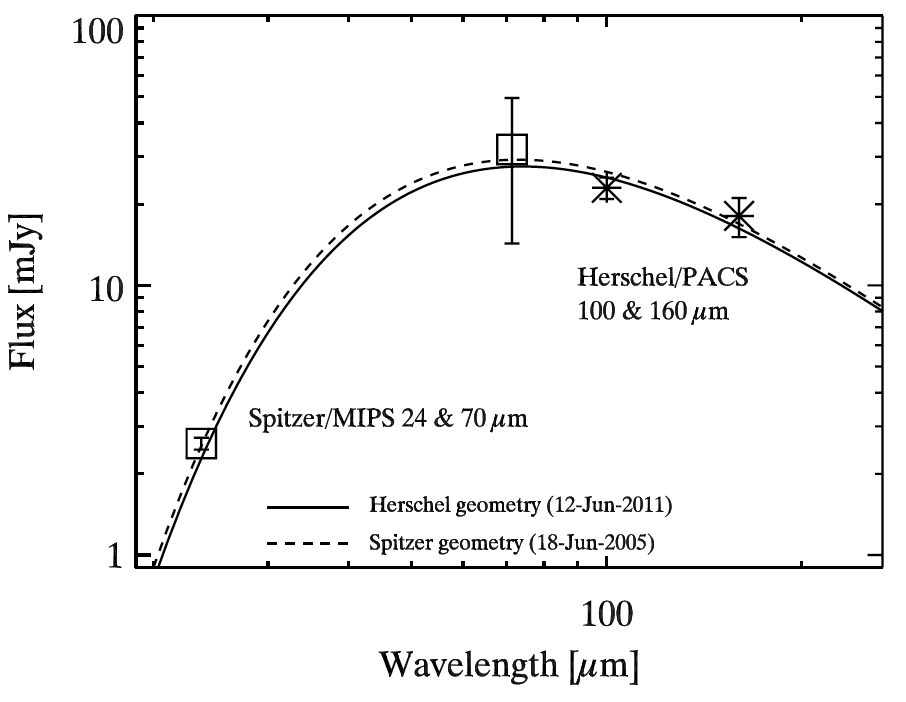}
%\centering\includegraphics[width=8cm,angle=90]{./sat_nereid_obsmod_other12.jpg}
\centering\includegraphics[width=8.1cm,angle=0]{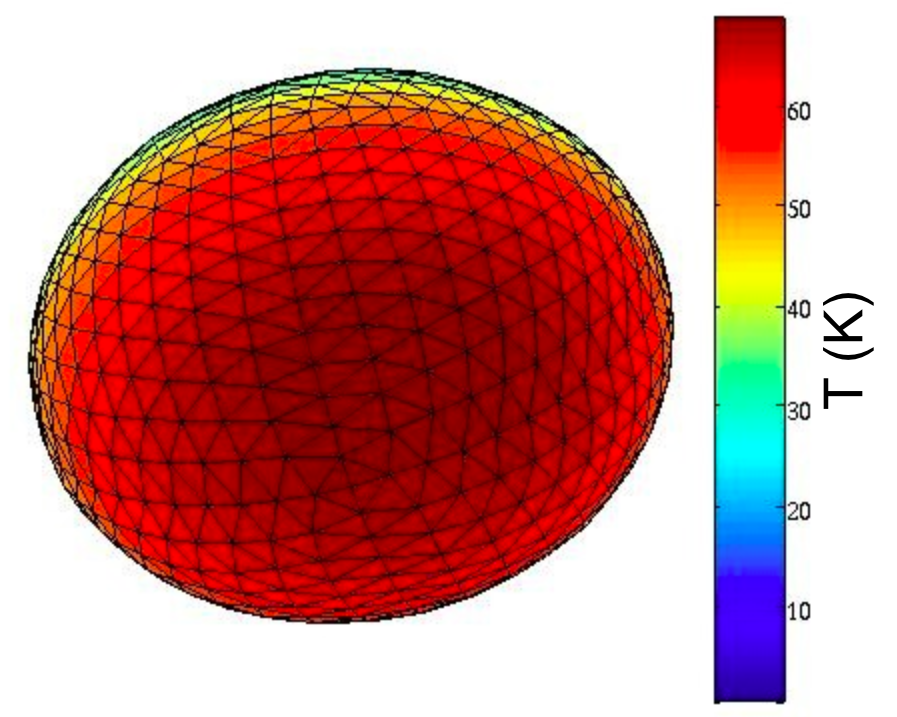}
\caption{Upper panel: Best fit TPM model of the thermal emission of
Nereid for an axis ratio parameter of X\,=\,0.0133 and a
spin axis orientation of $\lambda_p$\,=\,320\degr{} and 
$\beta_p$\,=\,32\degr{}. The solid and dashed lines correspond to the 
observing geometries at the Herschel/PACS and Spitzer/MIPS observation 
epochs, respectively. Lower panel: Temperature distribution of the
model above.}
\label{fig:tpmres}
\end{figure}
%%%%%%%%%%%%%%%%%%%%%

We also used a thermophysical model 
\citep[TPM, see][]{lagerros1996,lagerros1997,lagerros1998,muller1998,muller2002}. 
to characterise the thermal emission of Nereid. The TPM calculates the
temperature distribution on the surface of a body for a specific shape,
illumination, observing geometry, spin axis direction and rotation period. 
The model considers the thermal properties of the surface directly, 
including thermal inertia and surface roughness. 

We looked for solutions that matches the observed Spitzer/MIPS and 
Herschel/PACS far infrared fluxes the best, using the input fluxes
listed in Table~\ref{table:fluxes}. As in the case of the NEATM model,
we characterise the goodness of fit by calculating the reduced
$\chi^2$ values and we require that acceptable solutions should
fulfil the requirement of $\chi^2_r$\,$\le$\,$(1+\sigma)^2$. 

%One important parameter of the thermophysical models we use is surface roughness. 
%The TPM considers surface roughness with Gaussian random surfaces 
%\citep{lagerros1998} and we use the r.m.s. slope $\rho$ to characterise
%it. However, there is a straightforward transformation of $\rho$ 
%to e.g. hemispherical segments \citep[see][]{lagerros1998}, the
%other classical way surface features are used to be represented.   
%For the TPM models, we used a wide range of surface roughness 
%(\lele{0.1}{$\rho$}{0.9}) and thermal inertia values 
%(\lele{0.1}{$\Gamma$}{50\,\tiunit}).

Low ($\rho$\,=\,0.1) to intermediate ($\rho$\,=\,0.5) 
surface roughness cannot explain the observed fluxes as 
these values result in very high ($\chi^2_r$\,$\gg$\,3) reduced $\chi^2$. 
All acceptable solutions ($\chi^2_r$\,$<$\,1.7) are related
to "hot model settings", i.e., very high roughness 
($\rho$\,$\ge$\,0.7) combined with extremely low thermal inertia 
($\Gamma$\,$\ll$\,1\,\tiunit), except for configurations near to pole-on where 
hot temperatures are reached for a wider range of thermal inertias, as in 
these cases heat is not transported to the night side.
All acceptable settings ($\chi^2_r$\,$<$\,1.7) produce size and
  albedo solutions with D$_{eff}$=\,353--362\,km and 
  p$_V$\,=\,0.23--0.25. 
  
Assuming that Nereid has a thermal inertia in the range of
 1--5\,\tiunit, comparable to typical TNOs/Centaurs at similar distances
 \citep[see][]{Lellouch2013}, only model settings
  with a spin axis direction close to "pole-on", 
  \{$\lambda_p$,$\beta_p$\}\,=\,\{320$\pm$30\degr, 0$\pm$30\degr\}, produce acceptable
  flux predictions.
  The best-fit case of these models provided an effective size of D\,=\,357$\pm$13\,km 
  and a geometric albedo of p$_V$\,=\,0.24$\pm$0.02.
%* Figure "sat_nereid_obsmod_other11.ps" with SEDs and data
%  and obs/mod figure (rms=0.9, TI=3, D_eff=361.6, pV=0.236,
%  ObsEcLon/Lat of spin axis: 300.0, 0.0 = close to pole-on geometry) 
%Discussion:
%* I can produce acceptable fits to the observed fluxes whenever the
%  model settings lead to extremely high surface temperatures. This
%  is usually the case for pole-on, low TI, high roughness.
%* But also a spin axis close to "equator-on" could do that, but in
%  this cases I would need even more extreme roughness settings
%  (thinking about Hyperion which has a surface where probably even
%  extreme roughness settings in my TPM setup are not sufficient).
%* I played a bit with "equator-on cases" and extreme thermal settings
%  to produce high temperatures: I can easily produce chi2 < 1.7 fits,
%  but all solutions point towards extremely high roughness combined
%  with low TI. The radiometric sizes are then between 350-360 km,
%  albedos pV: 0.24/0.25.
%
%Conclusion on radiometric constraint on spin-axis orientation:
%1) very large roughness: close to pole-on configuration
%2) extremely large roughness: no constraint on spin-axis orientation
%   can be given.

Our TPM analysis above suggested a high subsolar latitude and a correspondingly low 
spin axis aspect angle. According to the spin-axis constraints 
discussed in Sect.~\ref{sect:spin} high $\beta_{ss}$ may occur for the highest possible 
axis ratio parameters only. This favours shape solutions of X\,$\approx$\,0.13, and a 
corresponding subsolar latitude of $\beta_{ss}$\,$\approx$60\degr{} 
(see Fig.~\ref{fig:vartheta}). We have chosen the highest possible shape
parameter value of X\,=\,0.133 (a:c\,=\,1.3:1) and the corresponding subsolar latitude of
$\beta_{ss}$\,=\,58\degr{}, and tested the feasibility of this configuration in a TPM
model. When high roughness ($\rho$\,=\,0.9) is assumed we obtained an acceptable 
$\chi^2_r$ value of 1.2, but this is associated with a very low thermal inertia of 
$\Gamma$\,=\,0.5\,\tiunit. For this solution the best fit effective diameter is D\,=\,335\,km,
with a correspondig geometric albedo of p$_V$\,=\,0.275. Application of an extreme roughness of 
$\rho$\,=\,1.0 provides very low reduced $\chi^2$ values of $\sim$0.5, even when the
"nominal" $\Gamma$\,=\,5\,\tiunit thermal inertia values are used. 
However, in these cases we obtain a somewhat smaller size of D\,=\,326\,km 
and a higher albedo of p$_V$\,=\,0.29. 
 
\section{Discussion}
\label{sect:Discussion}

The most important question about Nereid, as discussed in the literature,
was its shape and the precession behaviour of its spin axis. 
This was modelled in detail by 
\citet{Schaefer2008} and \citet{Hesselbrock2013}, with the main aim 
to explain the large amplitude brightness variations seen on different time
scales. These models assumed a long rotation period (72--144\,h) 
which was needed to achieve precession timescales of around a decade. 

One important finding of our paper is that we confirmed the short (11\fh594)
rotation period observed earlier by \citet{Grav2003} and \citet{Terai}. 
As it has been mentioned previously, a short rotation period 
implies a long spin axis precession time, as 
P$_{prec}$\,$\propto$\,P$_{orb}^2$/P$_{spin}$, where 
P$_{prec}$ is the precession period, P$_{orb}$ is the orbital 
period around Neptune, and P$_{spin}$ is the rotation period
\citep[see e.g.][]{Hesselbrock2013}. P$_{prec}$ values calculated from the 
11\fh594 rotation period are at least an order of 
magnitude longer than those assumed by \citet{Schaefer2008} 
and \citet{Hesselbrock2013}, even if an extremely elongated body is assumed. 
Therefore the spin axis orientation could not change significantly
in the last $\sim$15 years, and the precession of the spin axis cannot be 
the reason behind the large flux variations observed. 

In Sect.~\ref{sect:spin} we used this stability of the
spin axis orientation and obtained a maximum 
axis ratio of a:c\,=\,1.3 from the light curve amplitude analysis.
According to \citet{Hesselbrock2013}, forced precession 
of Nereid could be feasible only if the moon is considerably 
elongated with an $a$:$c$ axis ratio of $\sim$1.9:1. Such a high
axis ratio is excluded by our results, consistently with the long precession
timescales obtained. 

The shape solution favoured by the thermal analysis 
(axis ratio parameter of X\,=\,0.133) is also the one that 
provides the maximum possible light curve amplitude of $\Delta m$\,$\approx$\,0\fm13, 
with a peak in the 1960's (see Sect.~\ref{sect:spin}). 
Even this solution is unable to explain the extremely large brightness
variations observed some decades ago 
\citep[up to $\Delta m$ of 0\fm5, see fig.~4 in ][for a summary of these data]{Schaefer2008}. 
Whatever caused these variations, it cannot be the forced precession of a very 
elongated Nereid, as this scenario is inconsistent with our present data. 

The long rotation period assumed by \citet{Schaefer2008} and 
\citet{Hesselbrock2013} was partly based on the expectation
that if Nereid was formed around Neptune or captured at early times, then 
its original rotation -- which could have had a period of a few hours -- 
should have slowed down considerably, even if Nereid is just slightly elongated.
%(However, it could not slow down so much that it became tidally locked,
%otherwise there were no true
%rotational light curve, just changes with the orbital phase around Neptune. )
In this sense, Nereid rotates fast today -- this may favour a late 
capture (i.e. Nereid did not have enough time to slow down considerably) 
or indicate some other influence, like a collision, that may overwrite the 
rotational state. We note that \citet{Grav2004} suggested that Halimede, a small 
moon of Neptune, may be a fragment of Nereid broken off during a collision.  
This scenario for the origin of Halimede is supported by their 
similar colours and the relatively high probability of a collision between Nereid and 
Halimede in the timespan between the formation of the Solar System and today 
\citep{Holman2004}.

%%%%
%\begin{figure}
%\centering\includegraphics[width=8cm,angle=0]{./nereid_shape.png}
%\caption[]{A likely shape and orientation of Nereid nowadays, assuming
%a shape parameters of X\,=\,0.12 and a corresponding subsolar 
%latitude of $\beta_{ss}$\,$\approx$55\degr{} (see Sect.~\ref{sect:####}).  
%The spin axis aspect angle is $\vartheta$\,$\approx$\,35\degr. The body rotates 
%around its shortest $c$ axis; the direction of the Sun is indicated.}
%\label{fig:shape}
%\end{figure}
%%%%

As discussed in detail in \citet{Schaefer2008}, the Voyager-2 images
analysed by \citet{Thomas1991} could hardly be used to put constraints 
on the actual shape of Nereid, i.e. we did not have enough information 
to tell whether Nereid was spherical or notably non-spherical. 
The combination of the spin axis constrains and both the NEATM and TPM radiometry 
results favors a moderately elongated shape with a axis ratio parameter X\,$\approx$\,0.13, 
and correspondingly a present subsolar latitude of $\beta_{ss}$\,$\approx$\,60\degr
(a possible shape solution using X\,=\,0.133 is presented in Fig.~\ref{fig:tpmres}b). 
The moderately elongated shape put forward here for Nereid is feasible regarding 
observational constraints. However, for a moon of 
$\sim$350\,km in diameter a shape closer to spherical may be more likely
\citep[see e.g.][]{Schaefer2008} based on the shape information of giant 
planet moons of similar size. Most of these large ($>$100\,km in radius) moons are round 
to a few per cent (better than 10\%) accuracy. Our preferred solution for Nereid is 
further away from a perfect sphere. One exception among these satellites may be the 
Saturn moon, Hyperion, which is highly elongated despite that it is similar in 
size to Nereid (about 205$\times$130$\times$110\,km semi-axes). 
In the case of Hyperion the
elongated shape is explained by the high porosity of the interior ($\sim$40\%) 
that also leads to a 'sponge-like', high roughness surface 
where craters remain nearly unchanged over billions of years \citep{hyperion}. 
A porous internal structure and high roughness surface similar to that of 
Hyperion may explain well both the elongated shape and the radiometry analysis 
results obtained for Nereid in the present paper. 

\begin{figure}
\centering\includegraphics[width=8cm,angle=0]{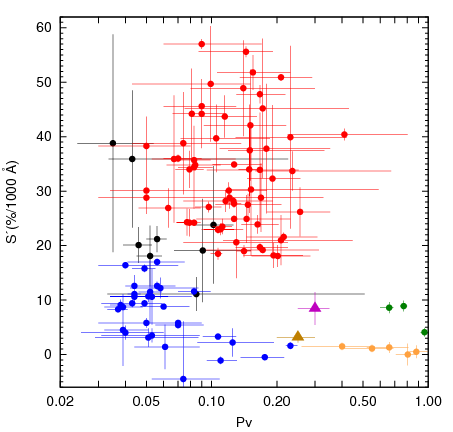}
\caption[]{Albedo-colour diagram of Centaurs and trans-Neptunian objects, 
also presenting the colours and albedos of Nereid 
\citep[brown triangle, data from this paper and][]{Grav2004} 
and Hyperion \citep[purple triangle,][]{Hicks2008,Thomas2010}. 
The colours and albedos of the other objects are taken from the original 
resources cited in \citet{Lacerda2014}. The colours of the symbols are
the same as in \citet{Lacerda2014}: red dots -- bright-red group, 
blue dots -- dark-neutral group, orange dots -- Haumea-type objects,
green-dots -- largest TNOs, {\bf black-dots -- objects with large 
uncertainties and ambiguous surface type}.}
\label{fig:albedocolour}
\end{figure}

In a recent paper \citet{Lacerda2014} presented an analysis 
of colours and albedos of Centaurs and trans-Neptunian objects, 
and identified two main groups whose existence can be considered
as an evidence for a compositional discontinuity in the young Solar System.
For comparison, we plotted the colour and albedo of Nereid in a diagram 
(Fig.~\ref{fig:albedocolour}) similar to fig.~2 in \citet{Lacerda2014},
presenting these dark-neutral and bright-red objects. Colours
are represented by spectral slopes, calculated in the same way as in 
\citet{Lacerda2014}. 
If Nereid was either formed around Neptune or captured during Neptun's outward
migration from the 20-30\,au distance range in the early Solar System, 
it is expected to exhibit a surface similar to the objects 
in the dark-neutral group. Indeed, Nereid is close in colour to the 
typical colours of objects in this group, but at the same time, 
its albedo is significantly larger than the typical value in this group, 
even larger than the albedos of most objects in the bright-red group. 
If Nereid was a member to the dark-neutral group, it would have the 
brightest surface among these objects. While Nereid is at the edge of the 
distributions of objects of both major groups in the albedo-colour plain, it 
is rather close in these characteristics to the Saturnian moon Hyperion
(purple triangle in Fig.~\ref{fig:albedocolour}). As mentioned earlier,
the surface roughness and internal structure of Nereid may as well be 
similar to those characteristic of this irregular satellite.  
  
\section{Summary}
\label{sect:summary}

In this paper we presented space born observation of Nereid, performed by the Kepler 
Space Telescope in the framework of the extended K2 mission, and results obtained 
from archival infrared data of the Spitzer Space Telescope and the Herschel Space 
Observatory. From the Kepler K2 data we obtained a light curve that was the same in 
period and similar in amplitude to those
obtained from ground based observations in 2001 and 2008. 
These observations together contrain the possible rotation states of Nereid very well. 
According to these results, Nereid is in a low amplitude apparent light curve state 
nowadays, but may have 
been in a much larger amplitude state some decades ago. We managed to exclude very 
elongated shapes with axis ratios above $1.3$:$1$; this also means that Nereid cannot be 
in a forced precession state due to tidal forces as it is not elongated enough for this 
process. This is a robust result as we assumed that the light curve of Nereid is solely 
caused by shape effects and albedo variegations existing on the surfce would just make 
the moon more spherical in this respect. 

We confirmed the size of Nereid obtained from Voyager-2 flyby data by an independent 
method -- radiometry based on infrared data. Both the NEATM and TPM thermal emission 
models resulted in similar effective size (D\,=335--345\,km) and albedo values 
(p$_V$\,=\,0.25--0.27). Both methods indicate
very high roughness (likely $\rho$\,$\approx$\,0.9) independently of shape, 
i.e. a surface with deep craters and very high surface coverage. 
Using the light curve and thermal emission results together
we obtain a likely moderately elongated shape (a:c\,$\approx$1.3:1) and a present 
spin axis aspect angle of $\vartheta$\,$\approx$\,30\degr{} for this irregular 
moon of Neptune. This shape may partly explain the larger light curves amplitudes 
observed some decades ago.

%Nereid is also similar to Hyperion concerning their albedos 
%(p$_V$\,=\,0.23 and 0.30 for Nereid and Hyperion, respectively). 
%Hyperion is 

\vspace*{2mm}

\section*{Acknowledgements}
%We thank the hospitality of the Veszpr\'em Regional Centre of the 
%Hungarian Academy of Sciences (VEAB) where most of our work was carried out. 
This project has been supported by the Lend\"ulet-2009 and 
LP2012-31 Young Researchers Program of the Hungarian Academy of Sciences, 
the Hungarian National Research, Development and Innovation Office
grants OTKA K-109276 and K-104607, NKFIH K-115709 and PD-116175, 
and by City of Szombathely under agreement no. S-11-1027. 
The research leading to these results has received funding from the 
European Community's Seventh Framework Programme (FP7/2007-2013) under 
grant agreements no. 269194 (IRSES/ASK), no. 312844 (SPACEINN), and the ESA 
PECS Contract Nos. 4000110889/14/NL/NDe and 4000109997/13/NL/KML. 
L.M. was supported by the J\'anos Bolyai Research Scholarship of the 
Hungarian Academy of Sciences.
Funding for the K2 spacecraft is provided by the NASA Science 
Mission directorate. 
The authors acknowledge the Kepler team for the extra efforts to 
allocate special pixel masks to track moving targets. 
All of the data presented in this paper were obtained from the 
Mikulski Archive for Space Telescopes (MAST). 
STScI is operated by the Association of Universities for Research in 
Astronomy, Inc., under NASA contract NAS5-26555. Support for MAST for 
non-HST data is provided by the NASA Office of Space Science via 
grant NNX13AC07G and by other grants and contracts. 
The useful comments of the referee are also appreciated. 

%\bibliographystyle{aj}
%\bibliography{K2asteroids}

{}

\end{document}